%% file: main.tex
\documentclass[a4paper, amsfonts, amssymb, amsmath, reprint, showkeys, nofootinbib, twoside, superscriptaddress]{revtex4-2}

\usepackage{titlesec}
\usepackage{ifthen}
\usepackage[english]{babel}
\usepackage[utf8]{inputenc}
\usepackage{newtxtext}
\usepackage{newtxmath}
\usepackage{physics}
\usepackage{siunitx}
\usepackage[colorinlistoftodos, color=green!40, prependcaption]{todonotes}
\input{preamble}

\usepackage[pdftex, pdftitle={Article}, pdfauthor={Author}]{hyperref} 

\usepackage{verbatim}
\usepackage{fancyvrb}
\newcommand{\vect}[1]{\boldsymbol{#1}}

\begin{document}

\title{Selection of prebiotic oligonucleotides by cyclic phase separation}

\author{Giacomo Bartolucci}
\thanks{Equal contribution}
\affiliation{Max Planck Institute for the Physics of Complex Systems, N\"{o}thnitzer Straße~38, 01187 Dresden, Germany}
\affiliation{Center for Systems Biology Dresden, Pfotenhauerstraße~108, 01307 Dresden, Germany}

\author{Adriana Cala\c{c}a Serr\~ao}
\thanks{Equal contribution}
\affiliation{Ludwigs-Maximilian-Universit\"{a}t M\"{u}nchen and Center for NanoScience,  Amalienstraße~54, 80799 Munich, Germany}

\author{Philipp Schwintek}
\thanks{Equal contribution}
\affiliation{Ludwigs-Maximilian-Universit\"{a}t M\"{u}nchen and Center for NanoScience,  Amalienstraße~54, 80799 Munich, Germany}

\author{Alexandra K\"{u}hnlein}
\affiliation{Ludwigs-Maximilian-Universit\"{a}t M\"{u}nchen and Center for NanoScience,  Amalienstraße~54, 80799 Munich, Germany}

\author{Yash Rana}
\affiliation{Max Planck Institute for the Physics of Complex Systems, N\"{o}thnitzer Straße~38, 01187 Dresden, Germany}
\affiliation{Center for Systems Biology Dresden, Pfotenhauerstraße~108, 01307 Dresden, Germany}

\author{Philipp Janto}
\affiliation{Ludwigs-Maximilian-Universit\"{a}t M\"{u}nchen and Center for NanoScience,  Amalienstraße~54, 80799 Munich, Germany}

\author{Dorothea Hofer}
\affiliation{Ludwigs-Maximilian-Universit\"{a}t M\"{u}nchen and Center for NanoScience,  Amalienstraße~54, 80799 Munich, Germany}

\author{Christof B. Mast}
\affiliation{Ludwigs-Maximilian-Universit\"{a}t M\"{u}nchen and Center for NanoScience,  Amalienstraße~54, 80799 Munich, Germany}

\author{Dieter Braun}
\email[ ]{dieter.braun@lmu.de}
\affiliation{Ludwigs-Maximilian-Universit\"{a}t M\"{u}nchen and Center for NanoScience,  Amalienstraße~54, 80799 Munich, Germany}

\author{Christoph A.\ Weber}
\email[ ]{christoph.weber@physik.uni-augsburg.de}
\affiliation{Faculty of Mathematics, Natural Sciences, and Materials Engineering: Institute of Physics, University of Augsburg, Universit\"atsstra{\ss}e\ 1, 86159 Augsburg, Germany}

\date{\today} 

\begin{abstract}
The emergence of functional oligonucleotides on early Earth required a molecular selection mechanism to screen for specific sequences with prebiotic functions. Cyclic processes such as daily temperature oscillations were ubiquitous in this environment and could trigger oligonucleotide phase separation. Here, we propose sequence selection based on phase separation cycles realized through sedimentation in a system subjected to the feeding of oligonucleotides.  
Using theory and experiments with DNA, we show sequence-specific enrichment in the sedimented dense phase, in particular of short 22-mer DNA sequences. The underlying mechanism selects for complementarity, as it enriches sequences that tightly interact in the condensed phase through base-pairing. Our mechanism also enables initially weakly biased pools to enhance their sequence bias or to replace the most abundant sequences as the cycles progress. 
Our findings provide an example of a selection mechanism that may have eased screening for the first auto-catalytic self-replicating oligonucleotides. 
\end{abstract}


\maketitle


\begin{figure}[!ht]
\includegraphics[width=1\columnwidth]{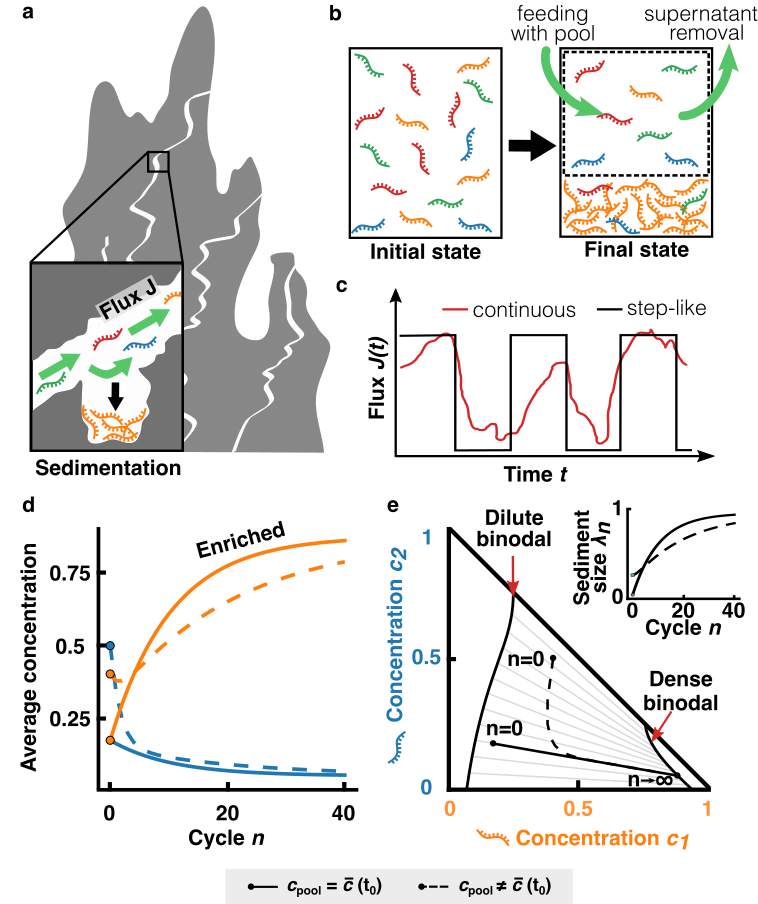}
\caption{
\textbf{Sequence selection via phase separation under varying feeding conditions.} 
\textbf{a} Illustration of a continuous oligonucleotide
flux ${J(t)}$ through porous rocks on early earth. By sedimentation, sequences that are capable of forming a condensed phase can enrich in pores, while others
are flushed away by the flow. 
\textbf{b} Cyclic version of \textbf{a}, where 
the feeding and removal steps are separated in time from the phase-separation kinetics. 
The dilute phase of the phase-separated pool is removed and the pool is replenished in a cyclic fashion.
\textbf{c} Comparison of the feeding flux ${J(t)}$
for continuous feeding \textbf{a} and cyclic feeding \textbf{b}. 
\textbf{d} Theoretical result for the selection kinetics in a system composed of solvent and two
sequences 1 and 2 with concentrations $c_1$ (orange) and $c_2$ (blue),
respectively. 
After each cycle, the system phase-separates and a fraction $\alpha$ of the dilute phase is removed and fed with a pool of fixed composition $\vect{c}_\text{pool}$; see Eq.~\eqref{eq:prot}. 
This pool can be equal to the initial average concentration ($\vect{c}_\text{pool}=\bar{\vect{c}}(t_0)$, solid lines) or a fixed pool that deviates from the initial pool ($\vect{c}_\text{pool}\not=\bar{\vect{c}}(t_0)$, dashed lines).
For these two cases, we chose $\alpha$=0.75 and $\alpha$=0.25, respectively.
Increasing the number of cycles $n$, sequence 1 gets enriched while sequence 2 becomes more diluted. 
\textbf{e} This selection kinetics can be represented as a trajectory in the corresponding phase diagram. 
During the kinetics, the condensed phase grows as indicated by an increase of its relative size $\lambda_n$ (inset).
}
\label{fig:origin}
\end{figure}

Oligonucleotides can catalyse biochemical reactions and store genetic information~\cite{Robertson2012, Kruger1982, Fedor2005}. The mechanisms through which functional oligonucleotides became sufficiently abundant are crucial to understanding the molecular origin of life~\cite{GilbertWalter1986}. In addition to sequence motifs, sufficient strand length is also a requirement for functional folds~\cite{Birikh1997}. Therefore, the assembly of long-chained prebiotic oligonucleotides has been the focus of many recent studies~\cite{Deck2011,Mariani2018,Walton2019,Wunnava2021}. 

However, the probability of randomly assembling a specific sequence of length $L$ with $m$ different nucleotides is proportional to $m^{-L}$. Since sequence space grows exponentially with sequence length, functional sequences are either not present or too dilute to interact and undergo chemical reactions. It is thus one of the central mysteries of the molecular origin of life how long enough sequences that enable self-replication could be selected from a large random pool of short non-functional oligonucleotides.

Due to the lack of complex biological machinery at the molecular origin of life, 
various physicochemical selective mechanisms have been considered~\cite{Budin2010}.
Examples are biased replication~\cite{Tkachenko2015,Tkachenko2018}, accumulation at surfaces due to gradients of temperature or salt~\cite{Mast2013}, and the accumulation at liquid-vapor interfaces~\cite{Morasch2019}. Another possible mechanism is related to the coexistence of two liquid-like phases. In particular, recent studies have shown that oligonucleotides condense and phase separate, forming coacervates~\cite{Aumiller2016,jeon2018}, liquid crystals~\cite{nakata2007end,Zanchetta2008} or hydrogels~\cite{Nguyen2017,xing2018}, which can lead to a local enrichment of specific oligonucleotides. 

An especially elegant case emerges when phase separation is caused directly by the base pairing of sequence segments among oligonucleotides. The strong interactions among complementary oligonucleotide strands (approximately $5 k_bT$ per base pair) lead to low saturation concentration above which phase separation occurs~\cite{Mitrea2016} and an oligonucleotide-dense phase that is composed of strongly correlated sequences. 
Thus, oligonucleotide phase separation via base pairing provides a mechanism to strongly accumulate a specific set of oligonucleotide sequences.

In a realistic prebiotic environment, 
such as an under-water rocky pore~\cite{Barge2017,Westall2018, Morasch2019},
the phase separation of oligonucleotides can be expected to be subject to periodic, typically daily, changes in the environment. 
In addition, such systems can exchange oligomers
with the environment via continuous fluxes, for example, driven by flow (Fig.~\ref{fig:origin}a).
Without phase separation, the oligomer composition approaches or remains at the composition of the environment, independent of oligonucleotide sequence.
However, when the oligomers can phase separate,
the oligomer-dense phase can grow by continuously recruiting sequences from the pool.
In this study, we ask whether this recruitment can significantly alter the oligonucleotide composition in the pore and thereby provide a physical mechanism of selection of specific sequences (Fig.~\ref{fig:origin}b).

We mimic the continuous exchange with the pool by a simplified cyclic protocol composed of two different steps: (i) A feeding step that corresponds to the replacements of the oligomer-poor phase by the pool, followed by 
(ii) a relaxation period toward phase separation equilibrium. 
In this case, the flux $J(t)$ follows a step-like profile in time; see Fig.~\ref{fig:origin}c. 
The key question is how much the oligonucleotide composition of the system can evolve away from the pool, which serves as a reference for the selection process. 
We investigate this question by theory and experiments using 
DNA. As a model oligomer, we decided to use DNA instead of RNA
since our study focuses on a selection mechanism that relies on base pairing, which is very
similar for both~\cite{santalucia1998unified}.
We show that phase separation combined with feeding cycles by a reference pool indeed provided a strong selection mechanism giving rise to distinct routes for molecular selection of specific oligonucleotide sequences.


\vspace{0.2cm}
\noindent \large{\textbf{Results and Discussion}} 
\normalsize
\vspace{0.1cm}

\noindent \textbf{Theory of cyclic phase separation with feeding.}
Here, we first discuss the theory governing the kinetics of an oligonucleotide mixture of volume $V_{}$ which is composed of $M$ different sequences.
This system undergoes cycles alternating between 
(i) a period where the material is exchanged by a pool and (ii) a period of phase separation (Fig.~\ref{fig:origin}b-c). Specifically,  within (ii),
the mixture phase separates into a dense and a dilute phase with sufficient time to relax to thermodynamic phase equilibrium, while
during the feeding step (i), a fraction $\alpha$ of the dilute phase is replaced by the pool.
After $n$ cycles, the average composition of the system is described by the $M$-dimensional vector, 
$\bar{\vect{c}}(t_{n}) =  
	\lambda(t_{n}) \vect{c}^\text{I}(t_{n}) + 
	\left( 1- \lambda(t_{n}) \right)\vect{c}^\text{II}(t_{n})$, 
where the vector components are the average concentrations of sequences. Moreover,  
$\vect{c}^\text{I}(t_{n})$  and $\vect{c}^\text{II}(t_{n})$ denote the concentrations of the dense and dilute phase, respectively, and $\lambda(t_n)=V^\text{I}(t_{n})/V_{}$ is the fraction of the system occupied by the dense phase, where $V^\text{I}(t_{n})$ denotes the volume of the dense phase.
The average composition of the system changes with cycle time $t_n$ due to the feeding step (i) and is given by (see SI, Sect.~\ref{app:cycles} for more information):
\begin{align} \nonumber
\bar{\vect{c}}(t_{n+1}) &=  \Big[
\big(1- \lambda(t_n) \big) \Big( \alpha \vect{c}_\text{pool}  + (1-\alpha)\vect{c}^\text{II}(t_{n}) \Big)\\
& \quad +\lambda(t_n) \vect{c}^\text{I}(t_{n})   \Big],
 \label{eq:prot}
\end{align}
where 
$\vect{c}_\text{pool}$ is the concentration vector characterizing the composition of the pool. 
The fraction of the dense phase $\lambda(t_n)$, and the concentrations of the dense and dilute phase $\vect{c}^\text{I}(t_{n})$  and $\vect{c}^\text{II}(t_{n})$  at cycle time $t_n$ can be determined 
by a Maxwell's construction in a $M$-dimensional state space for $\bar{\vect{c}}(t_n)$ obtained from solving the iteration Eq.~\eqref{eq:prot}.
The construction amounts to solving a set of non-linear equations 
that describe the balance of the chemical potentials and the osmotic pressure between the phases.
Their solution requires to estimate the sequence-specific interactions among the different oligonucleotides.
Details on the numerical method and the determination of interaction matrices are discussed in the SI, Sect.~\ref{app:hull}. 

To study the selection of oligonucleotide by cyclic phase separation, we considered the exchange of the oligonucleotide-poor phase by a pool of constant composition $\vect{c}_\text{pool}$, where the pool  acted as a reference for the selection kinetics. 
We studied two cases where the pool had the same composition as the 
initial average concentration at $t=t_0$, $\vect{c}_\text{pool}=\bar{\vect{c}}(t_0)$,
or both deviate from each other in terms of composition,
$\vect{c}_\text{pool} \not= \bar{\vect{c}}(t_0)$.
Representative time traces for both cases are shown in Fig.~\ref{fig:origin}d for a mixture composed of solvent and two oligonucleotide sequences.
We find that for both cases one sequence is enriched while the other sequence decreases in concentration as cycles proceed (orange and blue dashed lines, respectively). 
These concentration traces can be represented as trajectories of the average concentrations $\bar{\vect{c}}(t_{n})$ in the ternary phase diagram;  Fig~\ref{fig:origin}e.
Each point of the trajectory
that is between the dilute and dense binodal leads to a unique fraction of the dense phase, $\lambda(t_n)$, and a pair of concentrations corresponding to the dense and dilute phase, $\vect{c}^\text{I}(t_{n})$  and $\vect{c}^\text{II}(t_{n})$, respectively. 
As cycles proceed, the fraction of the dense phase $\lambda(t_n)$ grows (Fig~\ref{fig:origin}e, inset).
During this growth, sequence 1 is selected over sequence 2. As volume growth saturates at $\lambda(t_\infty)=1$, the selection process stops and the system settles in a stationary state. 

The sequence composition of this stationary state is set by the tie line defined by the pool composition $\vect{c}_\text{pool}$ (straight solid  black line in Fig~\ref{fig:origin}e).
Most importantly, the slope and length of this pool tie line determine if and how well sequences are selected. 
Only if the pool tie line deviates from the diagonal tie line in the phase diagram,
there is sequence selection during the growth of the dense phase.  
Selection is more pronounced if the pool tie line is longer in the phase diagram since more volume growth can occur. 
This case can be realized for example by strong interactions among sequences leading to the dilute and dense binodal branches being far apart in the phase diagram. 
Strikingly, both conditions, non-diagonal pool tie lines and strong sequence interactions, 
are particularly fulfilled in mixtures of oligonucleotides
that can interact via base pairing. 
From our theoretical study, we conclude that phase separation subject to cyclic feeding can provide a selection mechanism particularly relevant in 
oligonucleotides mixtures.

\begin{figure}[t!]
\includegraphics[width=1\columnwidth]{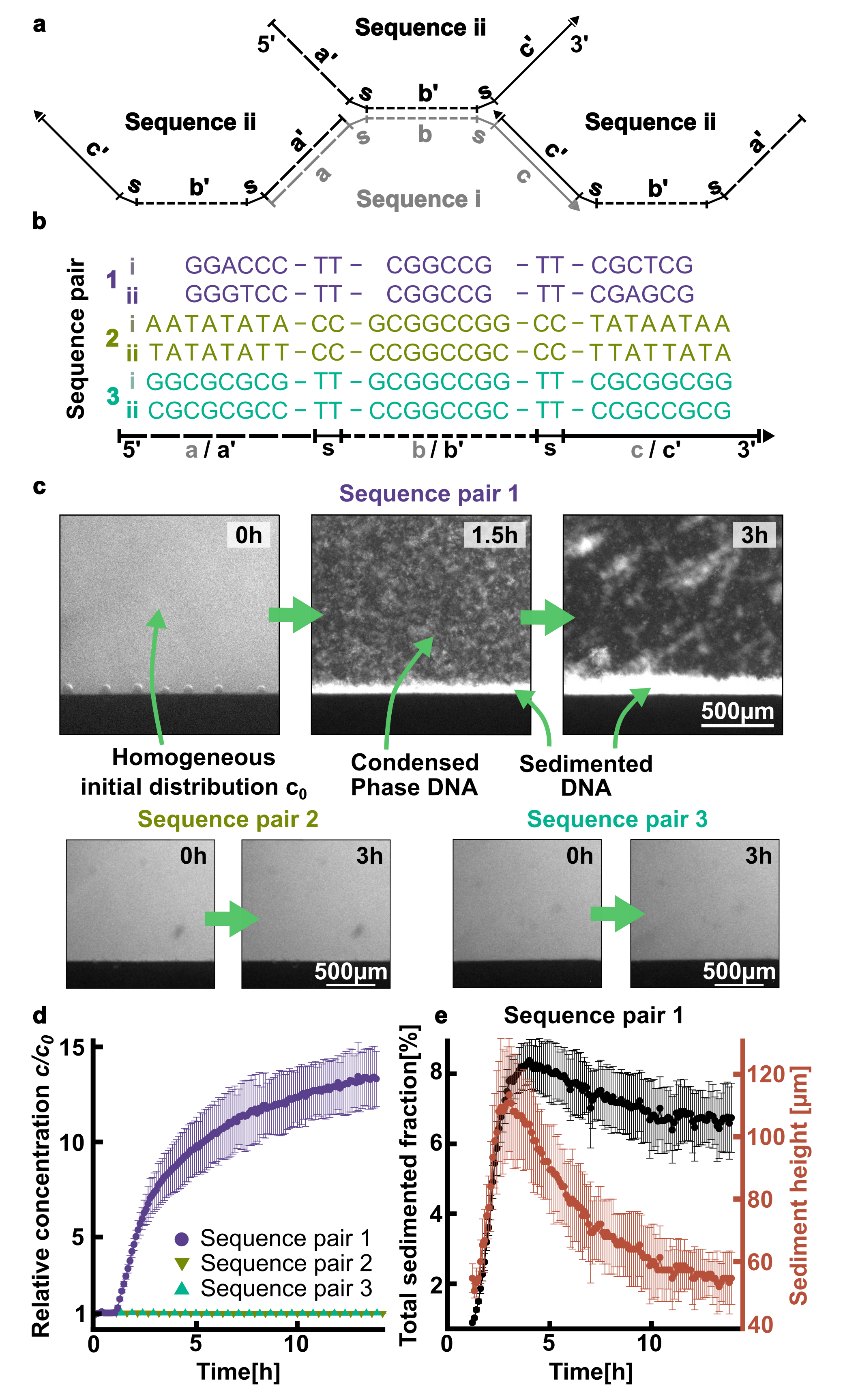}
\caption{\textbf{Condensation and sedimentation behavior of three sequence pairs.} 
\textbf{a,b} Sequence i is composed of three segments a, b, and c with spacer s. Its pair ii consists of reverse complements a’, b’ and c’. The inverted arrangement of a' and b' creates a network from four binding sites and prevents the formation of a linear double-stranded duplex.
\textbf{c} Fluorescence time laps images in a vertical, 500$\mu$m thin microfluidic chamber to prevent convection flow. Concentrations of strands were 25$\mu$M in a buffer of 10mM Tris-HCl pH 7, 10mM MgCl$_2$ and 125mM NaCl. Fluorescence labelling was provided by 5X Sybr Green I. After cooling from 65°C to 15°C, sequence pair 1 condensed into a solid phase and sedimented to the bottom of the chamber. Sequence pairs 2 and 3 showed a homogeneous fluorescence signal, not showing a condensed phase.
\textbf{d} Sequence pair 1 showed an up to 13-fold enhanced relative concentration while sequence pairs 2 and 3 showed no phase separation.
\textbf{e} The sedimentation behaviour of sequence pair 1 is studied by measuring its fluorescence. The total amount of sedimented DNA plateaued at 6-8$\%$ after 5h while the sedimented DNA contracted about 2-fold. The sticking of condensed DNA to the chamber walls could not be fully prevented. Error bars are standard deviations of three independent experiments.}
\label{fig:seq_design}
\end{figure}

\begin{figure*}[t] 
\includegraphics[width=1\textwidth]{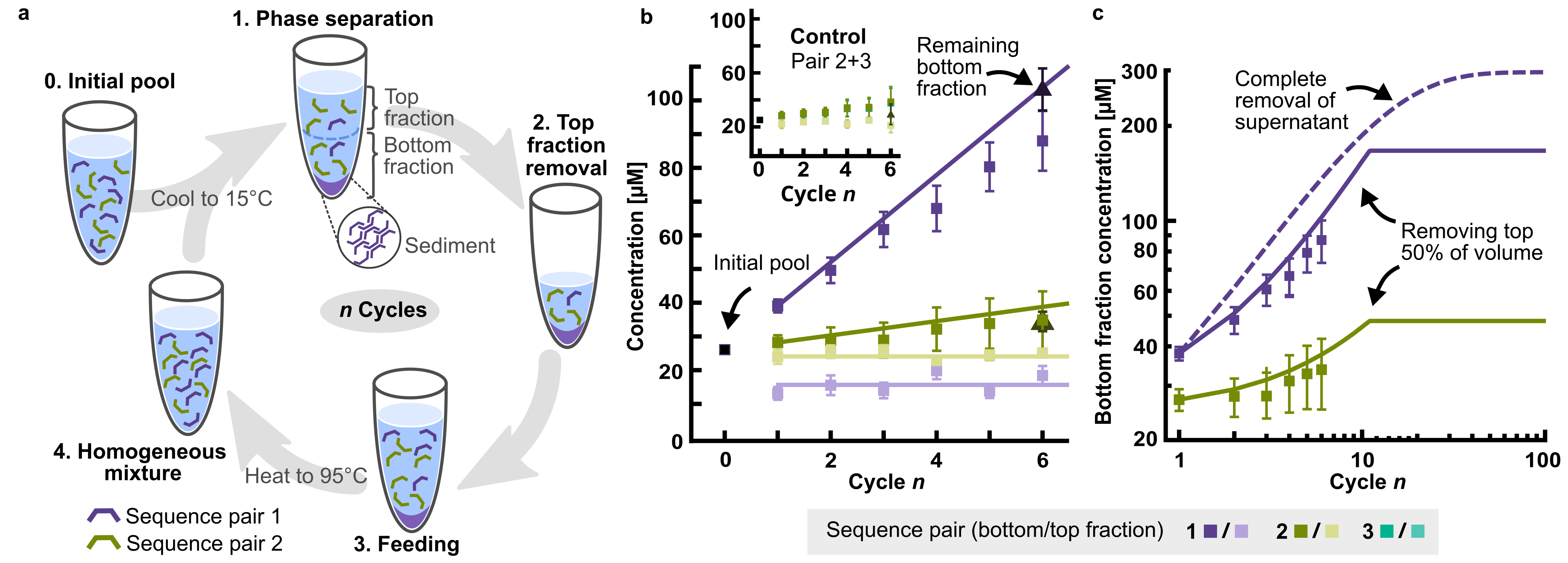}
\caption{ \textbf{Cycles of phase separation and feeding steps select specific oligonucleotide sequences from the initial pool.}
\textbf{a} Cyclic experimental protocol based on Fig.~\ref{fig:origin}b.\textbf{b} The initial pool contained a 25µM concentration of sequences pairs 1 and 2. After sedimentation, half of the top volume was removed ($\alpha$=0.5) and fed after each cycle with the same volume of the initial pool. Using quantification with HPLC, we found that sequence pair 1 (purple) was enriched while the concentration of sequence pair 2 (green) remained approximately constant. The same flat dynamics was found for a control system using non-condensing sequence pairs 2 and 3 (inset). In addition, the concentration of all supernatants and the final sediment were measured by absorbance at 260nm (triangular markers, see SI, Sect.~\ref{app:mass-balance}). \textbf{c} Solid lines show theoretical predictions. The bottom fraction concentration saturates once the sedimented DNA has filled  the bottom fraction of the chamber. If the whole supernatant could have been removed at each step of the cycle, only slightly amplified selection would be predicted (dashed line).}
\label{fig:exp_cycles}
\end{figure*}

\vspace{0.1cm}
\noindent \textbf{Observation of a condensed phase of DNA.} To experimentally scrutinize the selection propensity of phase separated oligonucleotide mixtures in the presence of feeding cycles, we have designed several experimental systems. The constructs were motivated by the theoretical model which suggested that a group of sequences were selected if they interacted strongly among themselves but weakly with other sequences. 
The sequence pairs (1-3) have three binding regions a, b and c which are separated by dimeric spacer sequences s. Each sequence pair had a complementary central sequence b and b'. The binding regions a/a' and c/c' are also complementary, but were reversed in their sequence such that three binding sites emerge while a self-binding hairpin or the formation of a fully complementary double strand could be avoided (Fig.~\ref{fig:seq_design} a-b). 
This partial base-paring between sequence segments enabled the formation of branched DNA aggregates with short sequences, leading to a condensed phase. In fact, mixtures of oligonucleotides were shown previously to be able to form condensed phases from oligonucleotides through mutual base-pairing of the individual strands~\cite{Nguyen2017,jeon2018, Morasch2019,saleh2020enzymatic, xing2018, xing2011self}, however required significantly longer sequences.

To characterize the phase separation propensity of our designed systems, we used time-lapse fluorescence microscopy (SI, Sect.~\ref{app:setup}). In particular, we imaged each system over time in temperature-controlled microfluidic chambers thin enough to prevent thermal convection. 
Prior to each experiment, the solutions were heated inside the microfluidic chamber to 65°C to ensure homogeneous initial conditions. 
The samples were then slowly cooled at a rate of  6K/min to 15°C and incubated at that temperature for at least $3h$. 
Microscope images for all three systems are shown in Fig.~\ref{fig:seq_design}c, where ``$0h$''  corresponds to the moment when the cooling step has reduced the temperature to 15°C. 
Within about $10min$, the first condensed phase DNA nucleated for sequence pair 1 (see SI, Sect.~\ref{app:videos}), grew within $1.5$ hours to a size of a few micrometers and sedimented at speeds of about $1mm/min$, forming a phase of sedimented DNA at the bottom of the chamber.
We could not observe a condensed phase for the sequence pairs 2 or 3 (see SI, Sect.~\ref{app:seq_2_3}).

The DNA concentration in the condensed phase increased up to 13-fold (Fig.~\ref{fig:seq_design}d). The total amount of molecules that sedimented saturates at about 8\% of the initial material at about 3 hours, decreasing then only slightly over time (Fig.~\ref{fig:seq_design}e, black data points). The height of sedimented DNA reached a maximum of about 100$\mu$m at 3h but then compacted to about half the height (Fig.~\ref{fig:seq_design}e, red data points). Similar behavior has been observed in literature for systems composed of longer DNA strands~\cite{Nguyen2017}.


\vspace{0.1cm}
\noindent \textbf{Cycles of condensation and feeding.} Based on our observation that sequence pair 1 can form a condensed DNA-rich phase, we experimentally scrutinize the theoretically proposed selection mechanism shown in Fig.~\ref{fig:origin}a-d that relied on a cyclic influx of material.
We subjected the phase-separating DNA to cyclic feeding steps by replacing the top fraction of the supernatant phase in the vial by the pool (Fig.~\ref{fig:exp_cycles}a, steps 1-3).
The theory suggests that exchanging the complete supernatant phase with the pool reached the final stationary state with minimal amount of cycles (see SI, Sect.~\ref{app:cycles}).
However, a complete removal of the supernatant phase by pipetting turns out to be experimentally difficult since this also risks to remove condensed DNA.
To avoid kinetically trapped states of condensing oligonucleotides, we additionally include annealing and melting steps in the cycle (Fig.~\ref{fig:exp_cycles}a, steps 3 to 4 and back to step 1). This procedure  enabled a fast relaxation to thermodynamic equilibrium after each feeding step. 

We investigated a system composed of equal fractions of the sequence pairs 1 and 2 (Fig.~\ref{fig:seq_design}b), where solely the sequence pair 1 showed phase separation before. 
As control we considered a system composed of sequence pairs 2 and 3 where we could not observe the formation of a condensed DNA phase (Fig.~\ref{fig:seq_design}b).
We determined the strand concentrations of each system in the top and bottom fractions of the vial using HPLC. We monitored the kinetics over six feeding cycles for the system composed of sequence pairs 1 and 2 (Fig.~\ref{fig:exp_cycles}b) and compare it to the non phase-separating control (inset in b). 
Both systems were intialized with equimolar concentrations of the two respective sequence pairs.

We found that the concentrations for the control hardly increased per cycles with slopes about or less than 2 \textmu M/cycle. 
This non-zero increase is probably due to the adhesion of strands to the vials surface.
For the phase-separating system with sequence pair 1 we observed that the concentration strongly increased, approximately linear with a slope of about (10.2 $\pm$  0.4) \textmu M/cycle (purple), while the sequence pair 2 in the mixture got only weakly enriched by the cycling. 
This observation confirmed that specific sequences could get selected by phase separation from the supernatant phase.  

In the experiments, the selection occurred concomitant to the growth of the condensed phase, which is consistent with our theoretical results. 
As cycles proceeded, the condensed DNA grew and lead to an increase of the concentration of the bottom fraction (Fig.~\ref{fig:exp_cycles}c).
In contrast, the concentration of the top fraction remained constant at about 14.7 \textmu M (Fig.~\ref{fig:exp_cycles}c, light purple). 
A constant supernatant concentration during cycles implies that system remained on the same tie line while the volume of the sedimented DNA was growing.
This corresponds to the simple theoretical scenario where the system is initialized at the pool tie line, as outlined in Fig.~\ref{fig:origin}d ($c(t_0)=c_\text{pool}$).

We quantitatively compared the experimental results for the bottom fraction concentration with the theoretical model. 
Since the experimental selection kinetics occurred on a single tie line, the sedimented and supernatant concentrations, $\vect{c}^\text{I}$ and $\vect{c}^\text{II}$, remained constant over time. For the supernatant concentration $\vect{c}^\text{II}$, we use the experimental concentration value of the top fraction. 
The sediment concentration $\vect{c}^\text{I}$ could be estimated for the theory using the experimental value for the initial average sequence concentration $\bar{\vect{c}}(t_{0})$ and the initial sedimented DNA size $\lambda(t_0)$.
Using these values, we find that the theoretical results (solid lines in Fig.~\ref{fig:exp_cycles}b) agree well with the experimental data points. 

Based on the agreement between experiment and theory, we could use the theory to extrapolate the selection kinetics for a larger amount of cycles (Fig.~\ref{fig:exp_cycles}c, solid lines).
For the experimental partial removal of the supernatant we found that selection approximately doubled after 20 cycles. The selection kinetics saturates because the condensed phase has grown to the volume corresponding to the bottom fraction. 
Finally, we used the theory to consider the ideal case of
a complete removal of the supernatant. We find that that for this ideal case the sequence pair can enrich by two-fold better than the for partial supernatant removal and more than 10 fold compared to the initial pool (Fig.~\ref{fig:exp_cycles}c, dashed lines).

In summary, we have shown experimentally that discrete cycles of feeding, i.e., replacing the supernatant with a reference pool, leads to the enrichment of specific sequence pairs based on the formation of condensed phases, confirming the theoretically proposed selection mechanism. 

\begin{figure}[!ht]
\includegraphics[width=1\columnwidth]{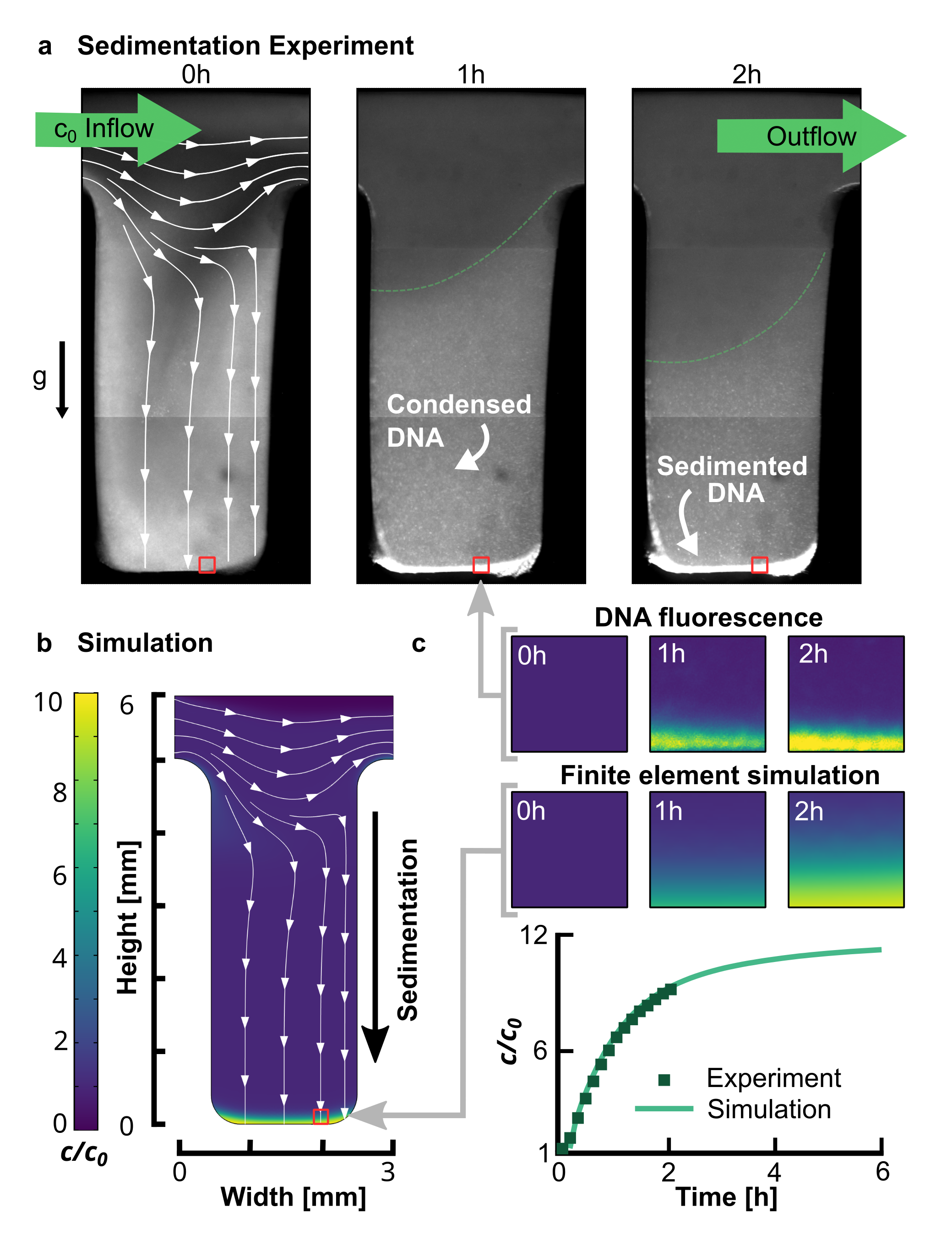}
\caption{
\textbf{DNA Phase separation under feeding with continuous flow. 
} 
\textbf{a} Fluorescence time-lapse images of a microfluidic chamber with a 2$\mu$m/s inflow of sequence pair 1. The condensed phase DNA sediments thus is not advected out of the chamber. This implements cycles of phase separation in a system with continuous flow. \textbf{b} Finite element simulation of fluid flow with sedimentation and diffusion confirms the experimental findings in detail when assuming a downward sedimentation speed of 0.1$\mu$m/s and a diffusion coefficient of 5$\mu m^2/s$. \textbf{c} The relative concentration profile match well between experiment and simulation, seen for a 200$\mu$m sized squared cut-out or when plotted over time.
}
\label{fig:flow}
\end{figure}


\vspace{0.1cm}
\noindent \textbf{Continuous DNA condensation in a feeding flow.}
A most elegant way to implement the cycle of Fig.~\ref{fig:exp_cycles} would be to create a continuous inflow of heated oligonucleotides that form a condensed phase upon cooling and sediment out of the feeding flow. Such a flow geometry mimics one of many possible environments on early Earth, such as a rocky pore of near a hydrothermal setting as illustrated in Fig.~\ref{fig:origin}a. To test the concept, we designed a microfluidic system with a continuous flow of the initial pool and compared it to a fluid flow theory. 
In this setup (see SI, Sect.~\ref{app:setup}), a pore (3 mm x 6 mm x 500$\mu$m) is connected to the feeding pool through a channel. If the flow is not too fast, the oligomers can phase separate and sediment and are not washed away towards the outlet, as shown by the fluorescent images of Fig.~\ref{fig:flow}a. In a prebiotic context, getting washed away would lead to further dilution into a larger reservoir such as the ocean. 
By the condensation of DNA, the sedimented DNA can accumulate and enrich inside the pore despite the continuous outflux. In our setup, we found that after 2 hours the concentration of sequence pair 1, $c(t)$, is enriched 8-fold relative to the inflow concentration $c_0$. 
The inflow and the choice of pore geometry has to be tuned not to perturb the sedimentation speed of the condensed phase DNA (Fig.~\ref{fig:flow}b). 
By numerically solving the hydrodynamic flow equations in addition to sedimentation and diffusion of condensed DNA, the simulated increase in concentration agrees well with the experiments in both space and time (Fig.~\ref{fig:flow}c).

\begin{figure}[!ht]
\includegraphics[width=1\columnwidth]{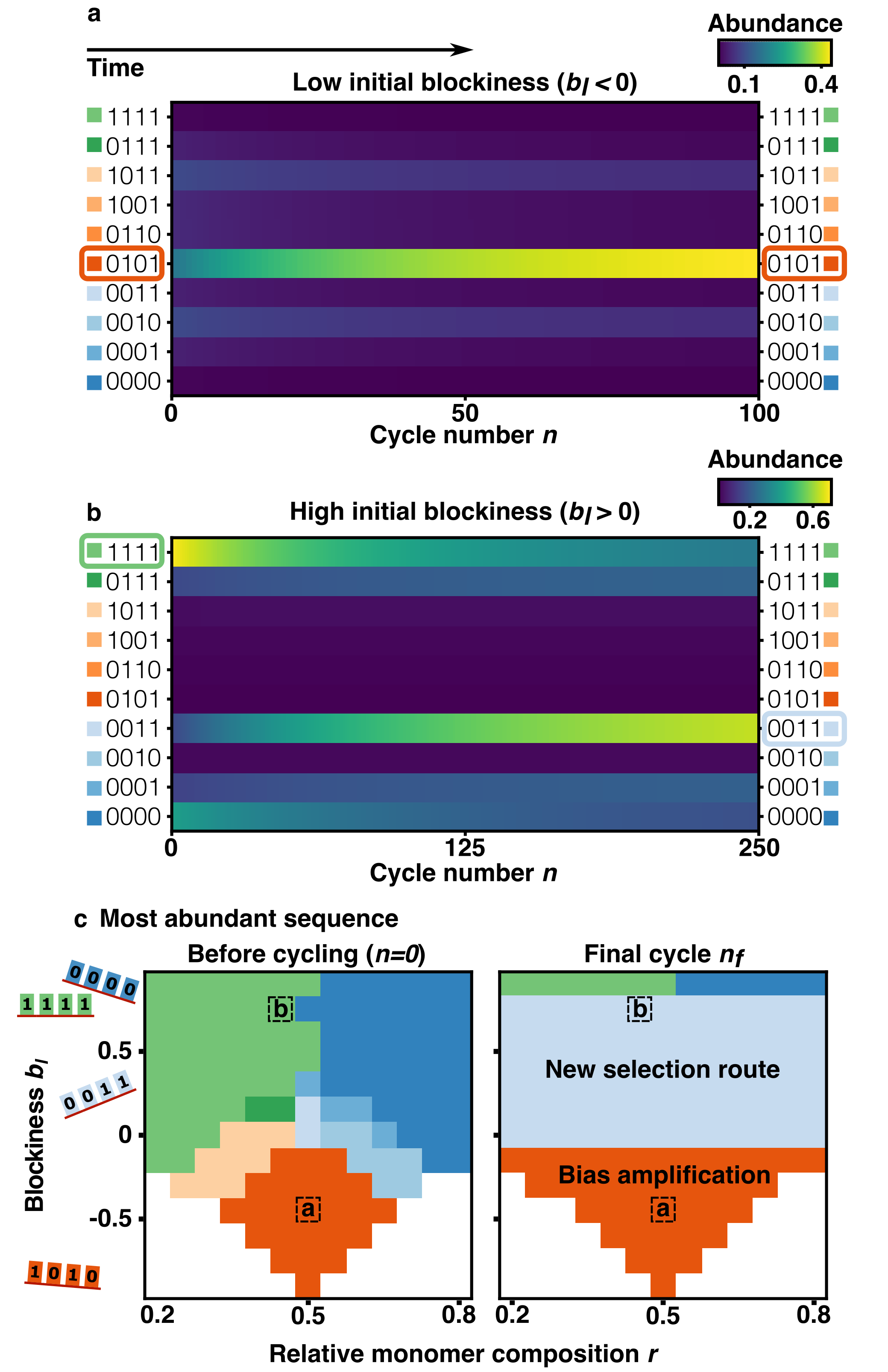}
\caption{\textbf{Cycles of phase separation can amplify pool bias or offer an alternative selection route}.
We consider pools composed of sequence ensembles characterized by two parameters: blockiness $b_l$ and relative composition $r$ of the monomers 0 and 1. 
\textbf{a,b} For low initial blockiness ($b_l<0$), the initially most abundant sequence (here: 0101) gets further enriched. In contrast,  for high initial blockiness ($b_l>0$), homopolymeric sequences (e.g. 1111 and 0000) are depleted with cycling while a more heteropolymeric sequence (0011) is strongly amplified. 
The color code indicates relative sequence abundances. 
\textbf{c} The most abundant sequence is shown for different values of $b_l$ and $r$ before cycling (left) after cycling $n_\text{f}$-times (right).
Further amplification of the initially most abundant sequence is found for low blockiness ($b_l<0$), while new selection routes can emerge at high blockiness ($b_l>0$).}
\label{fig:multi_theo}
\end{figure}

\vspace{0.1cm}
\noindent \textbf{Selection in pools with many sequences.}
Up to now, we have investigated specific pools composed of only a few designed sequence pairs for the proof of principle. 
It remains unclear how robust our proposed selection mechanism is for realistic pools that are formed via polymerization and contain many different sequences. 
To tackle this question,
we use our theory of cyclic oligonucleotide phase separation with discrete feeding cycles and consider pools 
that could emerge from the polymerization of different monomers.
For simplicity, we focus on sequences of fixed length $L$ composed of two different monomers, $0$ and $1$. 
Note that the monomers $0$ and $1$ could represent two different types of nucleotides or short segments with a specific nucleotide sequence.
Following Ref.~\cite{Fredrickson92}, sequence ensembles can be characterized by two parameters, the relative composition of the two monomers $r$ and the blockiness $b_l$. The latter determines the chain correlations of the two monomers: for $b_l=1$, the model favors homopolymers (..11.. or ..00..), while for $b_l=-1$, sequences are anti-correlated heteropolymers (..1010..);  see SI, Sect.~\ref{sec:polim_kin} for more details on the model.   
Phase separation of an ensemble of different sequences occurs once the system crosses the cloud point in the phase diagram; details see SI, Sect.~\ref{sec:cloud_pt}.
Subjecting such a phase-separated sequence pool to removal and feeding cycles, we find qualitatively different selection scenarios depending on the parameters $r$ and $b_l$. 
For initial pools of low blockiness ($b_l<0$), we find that the sequence bias of the pool is strongly amplified for a large numbers of cycles $n$ (Fig.~\ref{fig:multi_theo}a). 
This behavior results from the strong interaction propensity among sequences of the same type. 
Dominantly such sequences are recruited from the supernatant after pool replacement, while other sequences partition into the supernatant and get subsequently removed by the replacement step.
In contrast, 
for initial pools of high blockiness ($b_l>0$), the initial sequence bias is completely altered while cycling (Fig.~\ref{fig:multi_theo}b). 
In particular, homopolymeric sequences are disfavored, while heteropolymeric sequences get more favored as cycles proceed. The reason is that a homopolymeric sequences cannot interact with copies of itself, while 
more heteropolymeric sequences can. 
These trends are summarized for largely different sequence ensembles for different
values of monomer compositions $r$ and blockiness $b_l$ by
determining the most abundant sequence  before cycling and for large amount of cycles $n_\text{f}$ (Fig.~\ref{fig:multi_theo}c). 
We identify a domain at low blockiness ($b_l<0$) where the most abundant sequences are amplified, while new selection routes can emerge for sequence ensembles of intermediate or large blockiness ($b_l>0$). The later regime leads to a switch of the most abundant sequence when subjecting the system to a large number of feeding cycles. 

\vspace{0.2cm}
\noindent \large{\textbf{Conclusion}} 
\normalsize
\vspace{0.1cm}

\noindent Here we showed that the ability of oligonucleotides to form a condensed phases can give rise to an evolutionary selection mechanism if subject to feeding cycles. In particular,
replacing the supernatant phase with a constant pool composed of different 
oligonucleotide sequences
leads to growth of a condensed phase of specific sequences while others are depleted relative to the pool. 
We have quantitatively confirmed our theoretical predictions for discrete cycles by experiments using designed DNA sequences. 
In addition, we showed that the same mechanism also caused the selection of specific sequences in a prebiotic-relevant scenario where the system is subjected to a continuous flow of the pool.

The key property of the proposed selection mechanism is that it is highly sequence-specific, also in the presence of other interacting sequences.
Specifically, sequences that interact strongly with other sequences are enriched in the condensed phase while weakly interacting sequences are expelled from the condensate and thus leave the system through the removal step. 
A key observation of our work is that the selection mechanism also works well for very short oligonucleotides.
In our experiments, sequences of 22 nucleotides with base pairing regions of 6nt form cooperative base-pairing networks at room temperature and phase separate. 

Using the theory, we studied realistic, multi-sequence pools that result from polymerization. We found a robust and pronounced selection kinetics already for sequences that composed of only four segments of nucleotide sequences. We distinguish two qualitative  scenarios of sequence selection, where either the initial sequences bias is strongly amplified, or the initial bias is swapped, and other sequences are selected.

The robustness of our selection mechanism, particularly for short oligonucleotides, suggests its relevance at the molecular origin of life, where specific short-chained peptides, RNA, and DNA sequences were recruited during their assembly from prebiotic pools. 
The cyclic removal of weakly interacting sequences can guide the selection of longer sequences which face the molecule dilution by the exponentially growing size of sequence space. Moreover, the condensed phase could have provided enhanced stability against degrading chemical reactions such as catalytic cleavage~\cite{saleh2020enzymatic} or hydrolysis due to the duplex formation~\cite{Zhang2021}.
It is very interesting that the approach is confirmed by the recently found  correlation between catalytic sequences and phase separation in functional ribozyme polymerases~\cite{PhysRevLett.125.048104}.

\vspace{0.2cm}
\noindent \large{\textbf{Methods}} 
\footnotesize
\vspace{0.1cm}

\noindent \textbf{Strand design.} DNA oligonucleotide systems were designed using the NUPACK software package 3.2.2 (\cite{zadeh2011nupack}). The strands were constrained to contain three binding sites separated by spacers either composed of TT or CC. Systems that formed the intended secondary structure (each strand base-pairing with three other strands) were chosen (see SI, Sect.~\ref{app:Nupack}. The oligomers were ordered from biomers.net GmbH, in dry state, with high-performance liquid chromatography purification. The sequences were as follows (5'-3') - Sequence pair 1, sequence i: GGA CCC TTC GGC CGT TCG CTCG; sequence ii: GGG TCC TTC GGC CGT TCG AGCG; Sequence pair 2, sequence i: AAT ATA TAC CGC GGC CGG CCT ATA ATA A; sequence 2: TAT ATA TTC CCC GGC CGC CCT TAT TATA; Sequence pair 3, sequence i: GGC GCG CGT TGC GGC CGG TTC GCG GCGG; sequence ii: CGC GCG CCT TCC GGC CGC TTC CGC CGCG. All the strands stored at -20°C, diluted in nuclease free water at 200µM. Before every experiment, the strands were denatured at 95°C for 2min. 

\noindent \textbf{Reaction mixtures.} Initial pools of 15µL were prepared with 25 µM of each respective DNA strand, 10mM Tris Buffer-HCl pH 7, 5X SYBR Green I (intercalating dye; excitation 450-490nm, emission 510-530nm), 125mM NaCl and 10mM MgCl$_2$. The mixtures were heated to 95°C for 2 minutes to ensure full de-hybridization of the strands. The temperature protocol that allows hybridization and consequent phase separation, was \textit{i}. 95°C for 2min, \textit{ii}. 65°C for 10s, \textit{iii}. cooling to 15°C (ramp rate: 6K per minute), iv. 15°C for at least 3h. Temperature protocols were performed in a standard thermocycler (Bio-Rad CFX96 Real-Time System). Melting curves  were measured in triplicates using the same reaction mixture and temperature profile as for the sedimentation experiments (see SI, Sect.~\ref{app:melt}). Baseline correction using a reference measurement with only SYBR Green I. In the case of feeding cycle experiments, after sedimentation, 7.5µL of the supernatant, corresponding to 50\% of the initial volume, was removed by carefully pipetting only at the center of the meniscus to avoid removing material from the sediment. Afterwards, 7.5µL of the initial pool stock was added to the remaining bottom fraction. The aforementioned temperature protocol was then repeated, completing one feeding cycle. 

\noindent \textbf{Sedimentation imaging.}
The imaging experiments were performed in a microfluidic chamber containing multiple wells, cut out of 500 µm Teflon foil and sandwiched between two sapphire plates (see SI, Fig.~\ref{fig:wells}). The sample volume (about 15µL per well) was loaded by using microloader pipette tips. The temperature of the chamber was controlled using three Peltier elements. To remove the waste heat from the Peltier elements, a Julabo 300F waterbath (JULABO GmbH) was used to cool the back of the chamber. The entire chamber is held in place by screwing a steel frame on top using a homogenous torque of 0.2Nm. After loading, the wells were sealed with Parafilm to avoid evaporation. Monitoring of the sedimentation was performed using a self-built fluorescence microscope composed of a 490nm LED (M490L4, Thorlabs), a 2.5x Fluar objective (Zeiss) and the FITC/Cy5 H Dualband Filterset (AHF). Multiple wells could be imaged by moving the chamber perpendicularly to the light axis with two NEMA23 Stepper Motors and a C-Beam Linear Actuator (Ooznest Limited). Images were taken using a Stingray-F145B CCD camera (ALLIED Vision Technologies) connected via FireWire to a computer running a self-written Labview code operating camera, motors, LED’s and Peltier elements (see SI,  Fig.~\ref{fig:SISetup}). Flowthrough experiments were conducted using a similar chamber without Peltier Elements, only using the waterbath at homogeneous 15°C. In this case, the sapphires have holes on the backside, where an outlet and two inlet tubings were attached. Inlet tubing 1 contained 20uM of each strand of system 3 and 10X Sybr Green I, while inlet tubing 2 contained 20mM TRIS pH 7, 250mM NaCl and 20mM MgCl$_2$. Flowspeed was adjusted using the Nemesys Controler NEM-B002-02 D (Cetoni GmbH) with two 100µl syringes. Hardware was controlled using a self written labview (National Instruments) software (see SI,  Fig.~\ref{fig:flow}).

\noindent \textbf{High Performance Liquid Chromatography (HPLC).} Ion-pairing reverse phase HPLC experiments were carried out on a column liquid chromatography system equipped with an auto-sampler and a bio-inert quaternary pump (Agilent 1260 Infinity II Bio-Inert Pump G5654A, Agilent Technologies). A C18 capillary column (AdvanceBio Oligonucleotide 4.6x150 mm with particle size 2.7 µm, Agilent) was used to perform reverse phase liquid chromatography. The temperature of the autosampler was set to 4°C. The mobile phases consisted of two eluents. Eluent A was HPLC water (Sigma-Aldrich), 200mM 1,1,1,3,3,3, -Hexafluoro-2-propanol (HFIP) (Carl Roth GmbH), 8mM Triethylamine (TEA) (Carl Roth GmbH). Eluent B was a 50:50 (v/v) mixture of water and methanol (HPLC grade, Sigma Aldrich, Germany), 200 mM HFIP, 8 mM TEA. The injection volume for each measurement was 100 µL. The samples were eluted with a gradient of 1\% B  to 58.6\% B over the course of 45 minutes with a flowrate of 1mL/min. Prior to the gradient, the column was flushed with 1\% B for 5 minutes. Retention times were analyzed via a UV Diode Array Detector (Agilent 1260 Infinity II Diode Array Detector WR G7115A) at 260 nm with a bandwidth of 4nm. Samples were diluted for HPLC loading in the following manner: 7.5µL of sample, 105µL nuclease free water and 75µL of a 5M Urea solution. They were heated to 95°C for 2 minutes afterwards to ensure de-hybridization of the strands and dissolution of any sediment. Then, 105µL of the diluted samples were transferred into N9 glass vials (Macherey-Nagel GmbH) and stored at 4°C in the auto-sampler of the HPLC-MS system (1260 Infinity II, Agilent Technologies) until injection. 

\noindent \textbf{Finite-element simulations.} Simulations were performed in 2D using Comsol Multiphysics 5.4. The simulation file with all the detailed parameters is given in the supplement in binary format. Additionally, the simulation is given as an auto-generated report in a hierachical html compressed into a Zip-File. For more detailed information see SI, Sect.~\ref{app:melt}.


\vspace{0.2cm}
\noindent \large{\textbf{Data availability}} 
\footnotesize
\vspace{0.1cm}

\noindent The data that support the findings of this study are available from the corresponding author upon reasonable request.

\vspace{0.2cm}
\noindent \large{\textbf{Code availability}} 
\footnotesize
\vspace{0.1cm}

\noindent The code used to generate the data used in this study is available from the corresponding author upon reasonable request.

\let\oldaddcontentsline\addcontentsline
\renewcommand{\addcontentsline}[3]{}

\vspace{0.2cm}
\noindent \large{\textbf{References}} 
\footnotesize
\vspace{0.1cm}
\bibliography{References}

\let\addcontentsline\oldaddcontentsline

\vspace{0.2cm}
\noindent \large{\textbf{Acknowledgment}} 
\footnotesize
\vspace{0.1cm}
D.\ Braun acknowledges support from the European Research Council (ERC Evotrap, Grant Number 787356), the Simons Foundation (Grant Number 327125), the CRC 235 Emergence of Life (Project-ID 364653263), the Deutsche Forschungsgemeinschaft (DFG, German Research Foundation) under Germany´s Excellence Strategy – EXC-2094 – 390783311, and the Center for NanoScience (CeNS).
C.\ Weber acknowledges the European Research Council (ERC) under the European Union’s Horizon 2020 research and innovation programme (``Fuelled Life'' with Grant agreement No.\ 949021) for financial support.

\vspace{0.2cm}
\noindent \large{\textbf{Competing interests}} 
\footnotesize
\vspace{0.1cm}
The authors declare no competing interests.

\vspace{0.2cm}
\noindent \large{\textbf{Additional information}} 
\footnotesize
\vspace{0.1cm}
\noindent \textbf{Supplementary information} 
The online version contains supplementary material available at https://...

\newpage
\cleardoublepage

\onecolumngrid
\LARGE
\setcounter{page}{1}
\noindent \textbf{Supplementary information}
\normalsize

\author

\titleformat{\section}
  {\normalfont\large\bfseries}{Supplementary information \thesection:}{1em}{}
\renewcommand\thefigure{\thesection.\arabic{figure}} 
\renewcommand\thetable{\thesection.\arabic{table}} 
\setcounter{equation}{0}

\tableofcontents

\section{Multi-component phase separation subject to cyclic material exchanges}
\label{app:cycles}
\setcounter{figure}{0}
\setcounter{table}{0}

Here, we describe the concentration changes in a mixture of volume $V_{}$ that is composed of $M$ different oligonucleotide sequences subjected to periodic exchange of material. 
We introduce an $M$-dimensional vector $ \bar{\vect{c}}(t)$, where the components of this vector are concentrations of sequences. 
Starting from the initial state $ \bar{\vect{c}}(t_0)$, we perform $N$ exchange cycles, where each cycle is labeled with $n = 1, \dots, N$
and  composed of the two following steps:
\begin{itemize}
	\item[] \textit{Phase separation step}: The homogeneous mixture of concentration $\bar{\vect{c}}(t_n)$ phase-separates into two coexisting phases. We denote the  concentrations of the dense and supernatant phase as $\vect{c}^\text{I}(t_n)$ and $\vect{c}^\text{II}(t_n)$, respectively. The volume of the dense phase is $V^\text{I}(t_n)$ and thus the supernatant occupies the volume $(V_{}-V^\text{I}(t_n))$. Mass and particle numbers are conserved during the phase separation step: 
	\begin{equation}
	    \label{eq:cons}
	 \bar{\vect{c}}(t_n) = \left[\frac{V^\text{I}(t_n) }{V_{}} \vect{c}^\text{I}(t_n) + \frac{ V_{}-V^\text{I}(t_n) }{V_{}} \vect{c}^\text{II}(t_n) \right] \, .
	\end{equation}
	\\ 
	\item[] \textit{Partial supernatant removal step:} A constant fraction of the supernatant volume, $\alpha (V_{} -V(t_{n-1}))$, with the relative fraction $0<\alpha<1$, is replaced by the same volume that has the concentration of the pool, $\vect{c}_\text{pool}$.
	The average composition thus changes according to  
	\begin{align} \label{eq:app_protoi} \bar{\vect{c}}(t_{n+1}) &=  \left[\frac{V^\text{I}(t_n) }{V_{}} \vect{c}^\text{I}(t_{n}) + \frac{ V_{}-V^\text{I}(t_n) }{V_{}}  \Big( \alpha \vect{c}_\text{pool}(t_n)  + (1-\alpha)\vect{c}^\text{II}(t_{n}) \Big) \right]\, .
	\end{align}
\end{itemize}
For the general case where the initial average concentration is not equal to the average concentration of the pool,  $\bar{\vect{c}}(t_{0}) \not= \vect{c}_\text{pool}(t_n)$ (full lines in Fig.~\ref{fig:origin}d-e), we determine the phase compositions $\vect{c}^\text{I}(t_n)$ and $\vect{c}^\text{II}(t_n)$, and the phase volumes $V^\text{I}(t_n)$ at each cycle time $t_n$ by a construction is discussed in Sect.~\ref{app:hull}. 
Note that during he selection kinetics, the average concentration $\bar{\vect{c}}(t_{n})$ approaches the tie line corresponding to the pool.

For the special case where the initial average concentration is equal to one of the pool,
$\bar{\vect{c}}(t_{0}) = \vect{c}_\text{pool}(t_n)$ (dashed lines in Fig.~\ref{fig:origin}d-e), we can obtain an analytic solution, even for an arbitrary number of different components $M$. 
Only in this case,  $\vect{c}^\text{I}$ and $\vect{c}^\text{II}$, remain constant in time since the average concentration moves along the tie line defined by the pool. The iteration rule Eq.~\eqref{eq:cons}  simplifies to:
\begin{equation} 
\label{eq:rule} \bar{\vect{c}}(t_n) =  \bigg[ \lambda(t_n) \vect{c}^\text{I} + \left(1-\lambda(t_n)\right) \Big( \alpha c_\text{pool}  + (1-\alpha)\vect{c}^\text{II} \Big) \bigg] \,,
\end{equation}
where $\lambda(t_n)= {V^\text{I}(t_n)}/{V_{}}$ denotes the relative sediment volume. 
Using particle conservation for the pool concentration $\vect{c}_\text{pool}$ in Eq.~\eqref{eq:cons}, the interaction rule becomes:
\begin{align}
\label{eq:ruleRic}
	\bar{\vect{c}}(t_{n+1}) &= \Big[\alpha \lambda_0+ (1-\alpha \lambda_0 ) \lambda(t_n) \Big] \, \vect{c}^\text{I} +\Big[1 - \alpha \lambda_0- (1-\alpha \lambda_0 ) \lambda(t_n) \Big] \, \vect{c}^\text{II} \, ,
\end{align}
where  $\lambda_0=\lambda(t_0)$.
Using particle conservation at time step $t_{n+1}$,  
\begin{equation}
\label{eq:def}
	\bar{\vect{c}}(t_{n+1}) =  
	\lambda(t_{n+1}) \vect{c}^\text{I} + 
	\left( 1- \lambda(t_{n+1}) \right)\vect{c}^\text{II} \, , 
\end{equation}
we can identify the term in the first bracket of Eq.~\eqref{eq:ruleRic}, as $\lambda(t_{n+1})$, the size of the dense phase at $t_{n+1}$, and obtain a recursion relation:
\begin{equation}
\lambda(t_{n+1}) = a \, \lambda(t_{n}) + b \,,
\end{equation}
where  $a = 1-\alpha \lambda_0$ and $b = \alpha\lambda_0$. For large times, the system reaches a stationary state
\begin{equation}
\lambda(t_\infty)=\frac{b}{1-a}= 1 \, .
\end{equation}
This can be used to rewrite the recursion in terms of $\delta \lambda(t_n)$ with 
$\delta \lambda(t_n)= \lambda(t_\infty) - \lambda(t_n) = 1 - \lambda(t_n)$. 
As a result,
$\delta \lambda(t_{n+1}) = {a} \,\delta \lambda(t_n)$. Its solution reads $\delta \lambda(t_n) = \delta \lambda_0 \, {a}^n $ that can be written as 
\begin{equation}
\lambda(t_{n}) = 1- (1-\lambda_0) \left( 1- \alpha\lambda_0\right)^n  \,.
\end{equation}
This solution completely determines the evolution of the mean volume fraction:
\begin{align}
\label{eq:rule2b}
	\bar{\vect{c}}(t_{n}) &= \Big[ 1- (1-\lambda_0) \left( 1- \alpha\lambda_0\right)^n \Big] 
	\vect{c}^\text{I} +  (1-\lambda_0) \left( 1- \alpha\lambda_0\right)^n \vect{c}^\text{II} \, .
\end{align}
The characteristic number of iterations required to converge to the stationary state $\lambda(t_\infty) = 1 $ (and, consequently, $\bar{\vect{c}}_\infty = \vect{c}^\text{I}$) is
\begin{equation}
n_\text{c}= -\frac{1}{\log a} = -\frac{1}{\log [\alpha(1-\lambda_0)]} \, . 
\end{equation}

\section{Phase separation in multicomponent oligonucleotide mixtures}
\label{app:hull}
\setcounter{figure}{0}   
\setcounter{table}{0}

\begin{figure*}[h]
    \centering
    \includegraphics{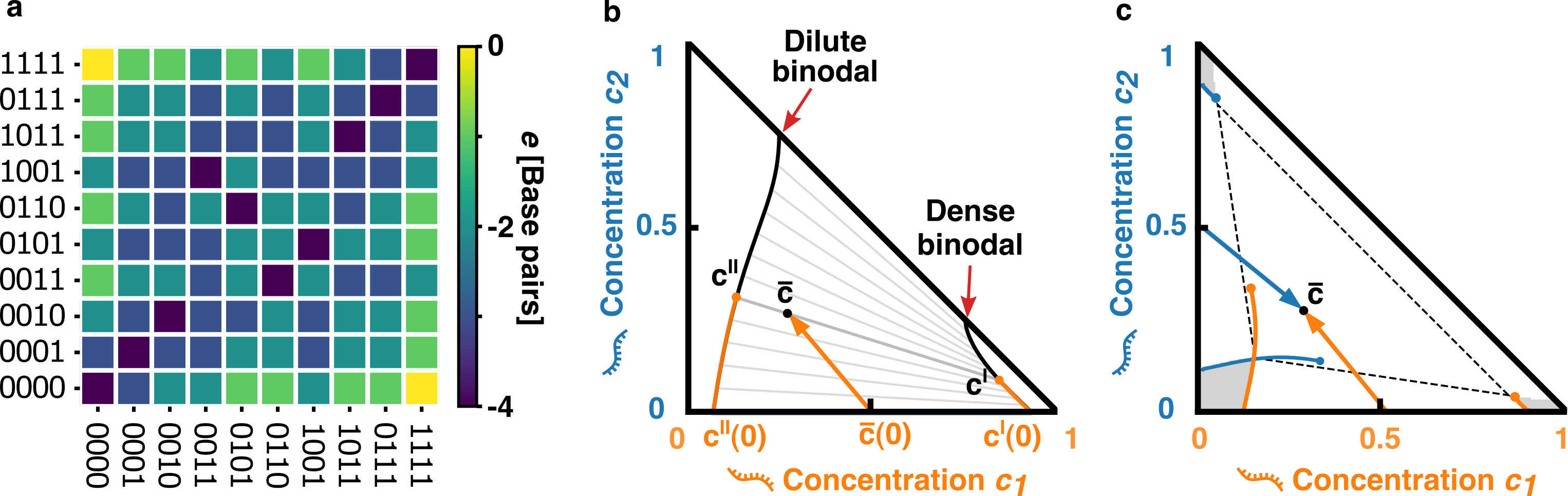}
    \caption{\textbf{Interaction matrix for short oligonucleotide sequences and phase diagram determination}. \textbf{a} Interaction matrix for sequences of fixed length $L=4$, which is calculated identifying the longest complementary part of the two sequences and then count the number of base pairs within that region. The negative sign indicate that sequences attract each other. \textbf{b} Determination of the phase concentrations and volume for a target average concentration $\bar{\vect{c}}$. The system is initialized at e.g.  the $c_1$-edge of the phase diagram and moving the average towards $\bar{\vect{c}}$ iteratively update previous results as guesses for the numerical solver. Here, $e_{11} = -k_B T$, $e_{12} = e_{22} -0.27 k_B T$ and $e_{ij} = 0$ otherwise, as in Fig.~\ref{fig:origin}.  \textbf{c} Identification of three phase coexistence regions as the locus of points $\bar{\vect{c}}$ for which the phase concentrations are different starting from different edges (orange and blue lines). Here $e_{11} =e_{22}= 5.33 k_B T$, and $e_{12} = 1.33 k_B T$, $e_{ij}=0$ otherwise.}
    \label{fig:SIHull}
\end{figure*}

To describe the phase behaviour of the system at each cycle, we chose the $T$-$V$-$N_i$ ensemble and introduce the free energy density $f\left(T,c_i\right)=F\left(T,V,N_i\right)/V$, where $F$ is the Helmholtz free energy depending on temperature $T$, volume $V$, and particle number $N_i$ of sequence $i$. 
The latter are related to concentrations via $c_i=N_i/V$.
Specifically, we use the following Flory-Huggins free energy density

\begin{equation}
\label{eq:f}
    f = 
    k_{B} T \left[
     \sum_{i=1}^{M+1} c_i \ln (c_i v) + c_\text{w} \ln (c_\text{w}  v_\text{w} ) + \frac{v^2}{2 v_\text{w} k_B T}\sum_{i,j = 1}^{M} e_{ij} c_i c_j \right]\, ,
\end{equation}
where $c_\text{w}$ and $v_\text{w}$ denote for the solvent concentration and molecular volume, respectively. We assume that all sequences have the saem molecular volume, $v_i = v$ for $i=1, \dots, M$. Furthermore, molecular volumes are are constant making the system incompressible. 
Thus, the solvent concentration can be expressed as:
\begin{equation}\label{eq:incompressibility}
c_\text{w} = \frac{1 - \sum_{i=1}^{M} c_i v}{v_\text{w}} \,.
\end{equation}
The parameters $e_{ij}$ encode the interactions energies among sequences. We estimated these parameters with the size of the maximum complementary portion of the two sequences, see Fig.~\ref{fig:SIHull}a. For simplicity, we neglect sequence orientation in the interaction matrix, i.e.,  $0101$ and $1010$ are considered the same). We also set to zero the interactions between sequences and the solvent $e_{i\text{w}} = 0$.

From the free energy density Eq.~\eqref{eq:f}, we can calculate the exchange chemical potentials of each sequence and the osmotic pressure:
\begin{subequations} \label{eq:mu_Pi}
\begin{align}
\bar{\mu}_i &= \frac{\partial f}{\partial c_i},  \qquad i = \, 1 \dots M  \, , \\
\Pi &= -f + \sum_{i=1}^M \bar{\mu}_i  c_i \, .
\end{align}
\end{subequations}
We searched for the domains in  concentration space where the system with average concentrations $\bar{c}_i$ demixes into a solvent poor and a solvent rich phases, where the dense and supernatant phase has the concentrations $c_i^\text{I}$ and  $c_i^\text{II}$, respectively. 
Number conservation relates both phase  concentrations to the average concentration 
\begin{align}
\label{eq:conc_constraint}
    \bar c_i = \frac{V^I}{V} c_i^\text{I}+ \frac{V-V^I}{V} c_i^\text{II} \, ,
\end{align}
where $V^I$ and $V$ are the volumes of the solvent poor phase and of the system, respectively.
The volume of the supernatant phase is $(V-V^I)$.
For $\bar c_i$ and $V$ fixed, the concentrations of the phases $c_i^\text{I}$ and $c_i^\text{II}$ and the volume $V^\text{I}$ are given by the solution of the following set of $(M+1)$ equations,
\begin{subequations} \label{eq:Max_constr}
\begin{align}
\bar{\mu}_i(c_i^\text{I}) &= \bar{\mu}_i(c_i^\text{II}) \, , \qquad  i= \, 1,2,\dots M \, , \\
\Pi(c_i^\text{I}) &= \Pi(c_i^\text{II}) \, ,
\end{align}
\end{subequations}
together with the $M$ constraints in Eq.~\eqref{eq:conc_constraint}. 
Finding the concentrations of the phases is in general a difficult task from a numerical perspective, in particular  for large numbers of sequences $M$ increases.
Numerical convergence requires on  suitable initial guesses. We thus developed a recursive method
to find the coexisting phases concentrations $\vect{c}^\text{I}$ and $\vect{c}^\text{II}$, and volume $V^I$ for a given average concentration $\bar c_i$. 
This method  is illustrated for a ternary mixture in Fig.~\ref{fig:SIHull}b.
The system is initialized at one edge of the phase diagram, corresponding to e.g. a system composed of sequence $1$ and solvent. In this subspace the set of equations~\eqref{eq:Max_constr} reduces to two and can be easily solved. We chose as initial average concentration $\bar{\vect{c}}(0) = (0.5,0)$ and  solve for $\vect{c}^\text{I}(0)$, $\vect{c}^\text{II}(0)$ and $V^I(0)$. We then consider a new average concentration $\bar{\vect{c}}(1)$ obtained displacing $\bar{\vect{c}}(0)$ along the direction $\bar{\vect{c}}(0)-\bar{\vect{c}}$ (orange arrow in Fig.~\ref{fig:SIHull}b). The magnitude of the displacement vector is fixed to $\Delta$ and must be chosen sufficiently small. Now we solve for $\vect{c}^\text{I}(1)$, $\vect{c}^\text{II}(1)$ and $V^I(1)$ corresponding to $\bar{\vect{c}}(1)$, using as initial guess for the solver $\vect{c}^\text{I}(0)$, $\vect{c}^\text{II}(0)$ and $V^I(0)$. We iterate this procedure, letting $\vect{c}^\text{I}(n)$, $\vect{c}^\text{II}(n)$ varying along the dense and dilute binodal branch, respectively (orange line) until the average reaches $\bar{\vect{c}}$. At this point, the phase concentrations have converged to $\vect{c}^\text{I}$ and $\vect{c}^\text{II}$ (orange dots along the dense and dilute binodal branch, respectively). 
We then start from the other edges of the phase diagram, corresponding to a system composed only of sequence $i$ and solvent, with $i=2, \dots M$.

Our proposed method  is able to determine volume and composition of two coexisting phases associated to a certain average concentration vector $\bar{\vect{c}}$, provided that there exists a segment joining $\bar{\vect{c}}$ and the midpoint of any edge that lies entirely in the demixing region (like the orange arrow in Fig.~\ref{fig:SIHull}b). In other words, any phase separating domain disconnected from all the corners would not be accessible.  
Another potential issue with this approach is that it accounts only for coexistence of two phases, neglecting states corresponding to three or higher coexisting phases. Luckily, regions corresponding to multi-phase coexistence can be self consistently detected with our algorithm. In fact, starting from different edges and solving Eqs. ~\eqref{eq:Max_constr} iteratively, would lead to different phase concentrations, if the average density lies within the multiphase coexistence region. This is exemplified in Fig.~\ref{fig:SIHull}c. Here, $\bar{\vect{c}}$ lies in the three phase coexistence. In this case starting from the $c_1$-only or $c_2$-only edge (orange and blue arrow, respectively) would lead to phase compositions that are different (orange and blue dots, respectively). Thus, in the following studies (see Section~\ref{app:kin_cloud}), we have checked that the average concentration $\bar{\vect{c}}$ lies in the two-phase coexistence region, making sure that starting from different corners we get the same phase concentrations.

\section{Polymerization kinetics and cloud point analysis for multi-component sequence pools}
\label{app:kin_cloud}
\setcounter{figure}{0}
\setcounter{table}{0}

We have applied the theory developed in sections~\ref{app:cycles} and~\ref{app:hull} to a system composed of all the possible sequences of length $L$ made by two different monomers $0$ and $1$. We focus on the protocol in which the initial sequence pool and the one used to replace the supernatant at each iteration are identical ($\vect{c}_\text{pool}=\bar{\vect{c}}(t_0)$). In this case, the evolution of the average concentration vector is governed by Eq.~\eqref{eq:prot}. According to this equation, the kinetics of $\bar{\vect{c}}$ is determined by the pool concentration $\vect{c}_\text{pool}$, the corresponding phase concentrations, $\vect{c}^\text{I}$ and $\vect{c}^\text{II}$, and the dense phase volume $V^I$ (see Appendix~\ref{app:hull}). 

We now discuss how assembly kinetics of monomers, which leads to the formation of polymers of length $L$, provides a natural choice of the pool composition, $\vect{c}_\text{pool}$, that is  crucial to determine the selection propensity of the mixture.

\begin{figure*}[h]
    \centering
    \includegraphics{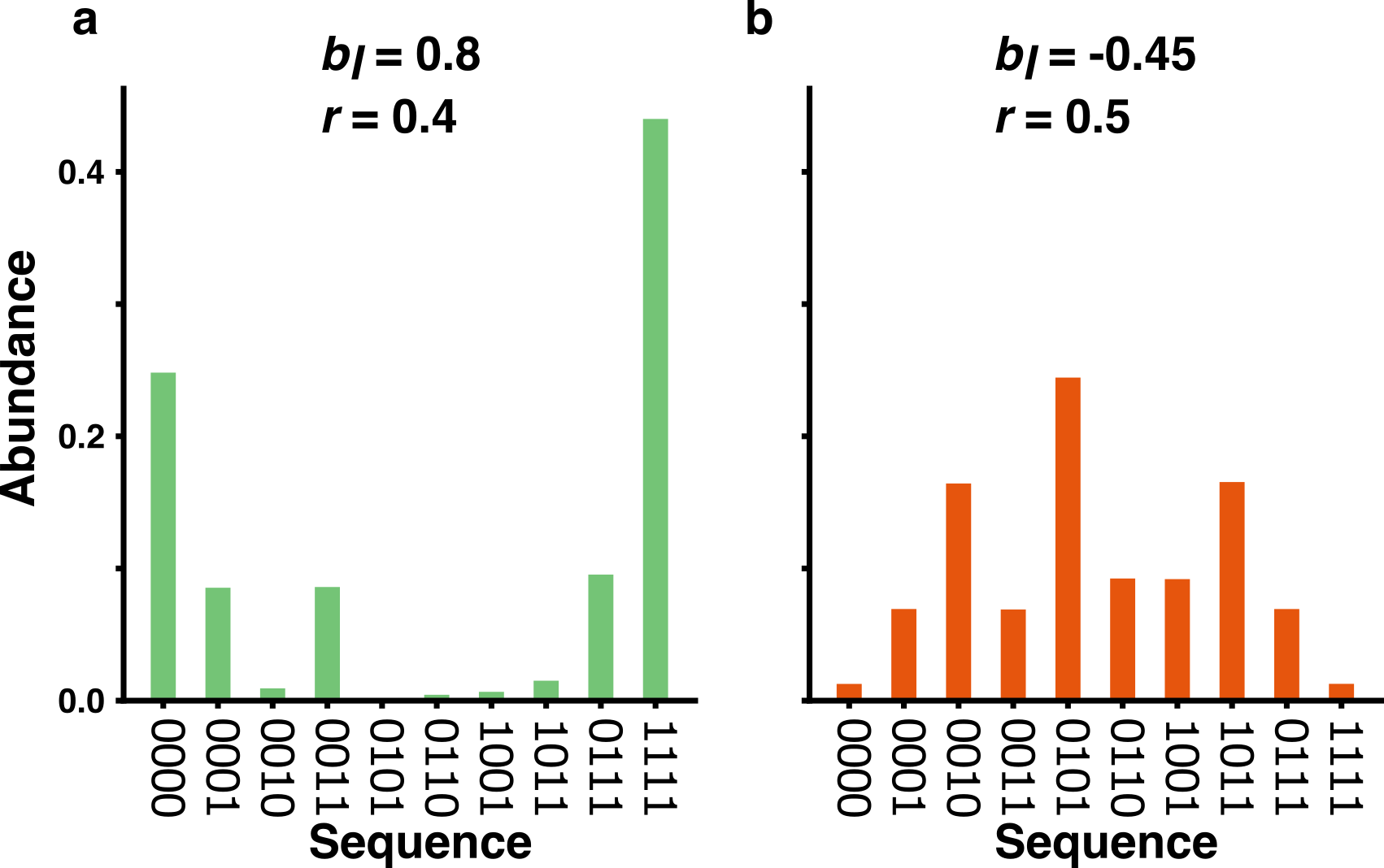}    \caption{\textbf{Sequence abundances generated via polymerization kinetics}. \textbf{a} $b_l = 0.8$, $r = 0.4$ \textbf{b} $b_l = -0.45$, $r = 0.5$. These are the same parameters as in Fig.~\ref{fig:multi_theo} a and b. Abundances are defined as sequence concentration normalized: $A_i = c_i/\sum_{i=1}^M c_i$.  In both cases $L =4$.} 
    \label{fig:SIExample}
\end{figure*}

\subsection{A minimal model for polymerization kinetics}
\label{sec:polim_kin}

To calculate the abundance of each sequence in a mixture, we use a simplified polymerization kinetics model that describes the assembly of monomers $0$ and $1$ into sequences of length $L=4$. This model was first introduced in Ref.~\cite{Fredrickson92}, and characterizes subsequent monomer addition using as input two parameters: $r$, the ratio between the number of $0$ monomers and the total number of monomers, and the blockiness $b_l$, quantifying the monomers correlation along the chain. There the authors show that, at steady state, $r$ and $b_l$ are in correspondence with the conditional probabilities $p_{ji}$, with $i,j = 0,1$, that a monomer of type $i$ is followed by a monomer of type $j$. In fact, we have

\begin{align}
\label{eq:blo_fra_def}
p_{00} &=  r (1-b_l)+ b_l \, ,\\
p_{11} &=  r (b_l-1)+ 1 \, ,\\
p_{10} &= 1-p_{00} \, ,\\
p_{01} &= 1-p_{11} \, .
\end{align}
For $b_l=1$,  monomers of the same type are neighbored leading to only two sequences, i.e., $0000$ and $1111$. In contrast, 
for $b_l=0$ there are no correlations and different monomers appear randomly. For $b_l=-1$, different monomers alternate generating a single sequence, i.e., $0101$ for $r = 0.5$. 

We can now fix $r$ and $b_l$, calculate $p_{ji}$ and use them to construct a long chain of length $L_\text{c} $. The sequence pool is then obtained by chopping the long chain into sequences of length $L$. In Fig.~\ref{fig:SIExample}, we show an example of the abundance of each sequence for different choices of $r$ and $b_l$. Abundances are defined as sequence concentration normalized without taking the solvent into account: 
\begin{align}
\label{eq:abu}
    A_i = \frac{c_i}{\sum_{i=1}^M c_i} \, .
\end{align}
For simplicity we neglect sequence orientation in the interaction matrix (see Fig.\ref{fig:SIHull}a) and thus  grouped sequences which are related via reflection along the mid plane (e.g., $0101$ and $1010$). 
\subsection{Cloud point determination}
\label{sec:cloud_pt}

\begin{figure*}[h]
    \centering
    \includegraphics[width=1\textwidth]{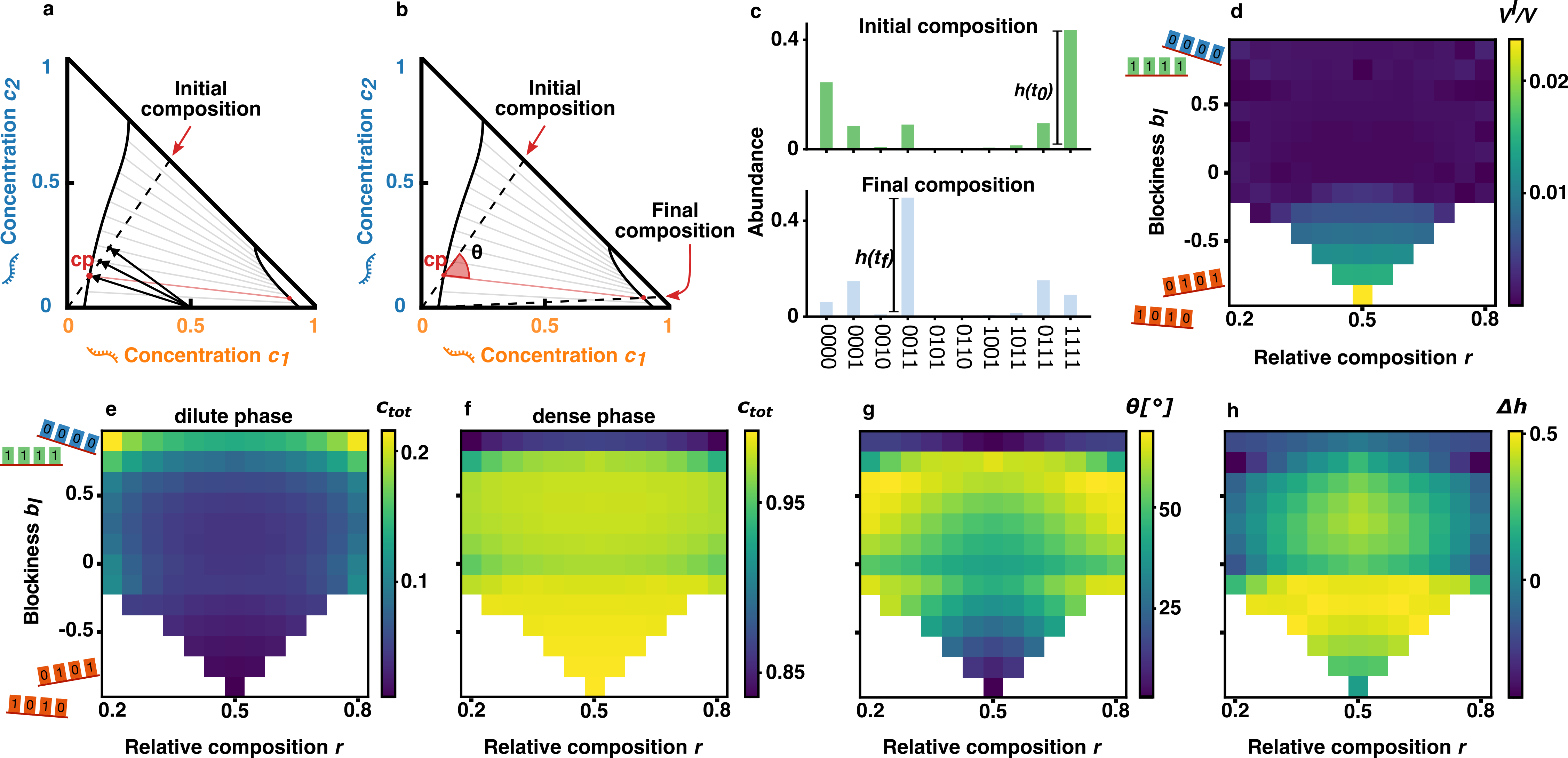}
    \caption{\textbf{Identification and characterisation of the cloud point in a multi-component mixture.} \textbf{a, b}  We illustrate the location of the cloud point in the case of a binary mixture with $e_{11} = -k_B T$, $e_{12} = e_{22} -0.27 k_B T$ and $e_{ij} = 0$ otherwise, as in Fig~\ref{fig:origin}e. \textbf{a} Once the relative sequence concentrations are fixed, the cloud point (red dot with ``cp" label) can be located by targeting points with increasing solvent amount and applying the method described in Fig.~\ref{fig:SIHull}. \textbf{b} The angle $\theta$ measures the distance in composition between the dense and dilute phases which, with our choice of protocol, are the final and the initial state of the mixture. In \textbf{c}-\textbf{h} we focus on the case of $L=4$ sequences, with the interaction matrix depicted in Fig.~\ref{fig:SIHull} and initial composition obtained varying $r$ and $b_l$ as explained in the previous Sec.~\ref{sec:polim_kin}. Deviating from the initial composition (high $\theta$), does not always correspond to selection. In fact, to assess whether the final pool has a stronger sequence enrichment than the initial pool, we  compare the initial abundances with the final and maximal abundances, indicated in \textbf{c} as $h(t_f)$ and $h(t_0)$. Here $r = 0.4$ and $b_l = 0.8$, as in Fig.~\ref{fig:multi_theo}). In \textbf{d} we check that the cloud point is efficiently located, i.e., the initial pool is characterised by $V^\text{I}/V \ll 1$. In \textbf{e} and \textbf{f} we show the total concentration at the cloud point of the dilute and dense phase, respectively. \textbf{g} The angle $\theta$ reveals that the final pool deviates significantly from the initial composition for positive, intermediate blockiness values, i.e., $b_l\simeq 0.5$. \textbf{h}  $\Delta h$ shows that the final pool has improved in selectivity with respect to the initial pool negative, moderate blockiness values, around $b_l\simeq -0.5$. Here, $k_B T = 0.8$ in units of a single base pair energy.}
    \label{fig:SICloud}
\end{figure*}

For each $r$ and $b_l$, we determined the minimum total oligomer concentration $c_\text{tot} = \sum_{i=1}^{M} c_i $ that leads to phase separation. This corresponds to locating the cloud point in the phase diagram, defined as the interception between the binodal and the line in which the sequence abundances are constant. The cloud point is of  experimental relevance since low oligomer concentrations are practically accessible. The procedure is illustrated in Fig.~\ref{fig:SICloud}a for a ternary mixture. Here, $e_{11} = -k_B T$, $e_{12} = e_{22} -0.27 k_B T$ and $e_{ij} = 0$ otherwise, as in Fig~\ref{fig:origin} e. The black solid line represents the binodal, while the dashed line is the equi-composition line which corresponds to a constant $c_1/c_2$ ratio. The latter is the two-dimensional analogue of the line defined by fixing the sequence abundances and varying the total oligomer concentration that correspond to a fixed choice of $r$ and $b_l$. The cloud point is located at the interception between the two lines (red dot indicated with ``cp'' in Fig.~\ref{fig:SICloud}a). To locate the cloud point, we repeat the procedure described in Appendix~\ref{app:hull} targeting average concentrations  on the equi-composition line with decreasing total oligomer concentration $c_\text{tot}$. 
In red we highlight the cloud point tie line. Its interception with the binodal gives the dense phase associated with the cloud point, see Fig.~\ref{fig:SICloud}b. The latter  determines the final system composition, since we chose a point along the tie line which also correspond to the average composition of the pool. An example of initial and final sequence abundances, corresponding to $r=0.4$ and  $b_l=0.8$ (same values as in Fig.~\ref{fig:multi_theo}b, in the main text), is displayed in Fig.~\ref{fig:SICloud}c.    

With this method, we compute the point along the fixed-composition line which belongs to the demixing region but is the closest to the cloud point. We compute the corresponding relative dense phase size $\lambda=V^\text{I}/V$, see Fig.~\ref{fig:SICloud}d, which is always lower than a few percent. This confirms that the distance between the average density obtained with this method is very close to the dilute phase vector, meaning that we can efficiently locate the could point. In Fig.~\ref{fig:SICloud}d,f we show the total oligomer concentration $c_\text{tot}$ at the cloud point and in the corresponding dense phase. As the blockiness decreases, $c_\text{tot}$ in the dilute phase and the dense phase respectively decrease and increase, in a non-monotonic fashion. 

As anticipated, we chose the pool along the cloud point tie line (very close to the cloud point itself, just inside the demixing region). To characterize selection propensity, we introduce the angle between the equi-composition line and the pool tie line (depicted in Fig.~\ref{fig:SICloud}b) 
\begin{align}
    \label{eq:theta_param}
    \theta =\arccos \left( \frac{\vect{c}^\text{II} \cdot \left( \vect{c}^\text{I} -\vect{c}^\text{II} \right)}{ \lVert\vect{c}^\text{II} \rVert  \, \lVert \vect{c}^\text{I} -\vect{c}^\text{II}\rVert }\right) \,.
\end{align}
This parameter quantifies how much the final state deviates from the initial pool. 

We then introduce the maximum abundance and its variation at the end of the protocols cycles 
\begin{align}
    \label{eq:h_param}
    h(t) = \max_{i = 1, \dots M} A_i(t) \qquad \text{and} \quad \Delta h = h(t_f) - h(t_0) \, ,
\end{align}
where $A_i$ are the sequence abundances, see Eq.~\eqref{eq:abu} and graphical representation in Fig.~\ref{fig:SICloud}c. Positive, large values of $\Delta h $ represent cases in which the final sequence distribution is more peaked around one single sequence, as opposed to the initial distribution, hence strong selection propensity. In Fig.~\ref{fig:SICloud}g and h we show how these quantifiers vary as a function of $r$ and  $b_l$. Pools characterized by slightly negative blockiness values have strong selection propensity, but the sequence distribution is similar before and after phase separation cycles (high $\Delta h $ but low $\theta $). This means that the initial distribution is already dominated by a sequence ($0101$) and after cycles of phase separation the same sequence remains the most abundant, and becomes even more dominant. This is the scenario that we called ``bias amplification" in Fig.~\ref{fig:multi_theo}c. Pools characterized by intermediate blockiness values show weaker selection propensity, but as $r$ deviates from $1/2$, the initial and final sequence distributions are different (moderate $\Delta h $ but high $\theta$). In this case the initial distribution is biased towards a particular sequence ($0000$ or $1111$, for example) and after cycles of phase separation the most abundant sequence is another one ($0011$) which is slightly more dominant than the initial most abundant sequence. This case corresponds to the scenario of a ``new selection route", see Fig.~\ref{fig:multi_theo}c. Note that in Fig.~\ref{fig:SICloud} and in Fig.~\ref{fig:multi_theo} in the main text, $k_BT = 0.8$ in units of a single base pair energy.



\section{NUPACK folding predictions}
\label{app:Nupack}
\setcounter{figure}{0}
\setcounter{table}{0}

\begin{figure*}[h]
    \centering
    \includegraphics{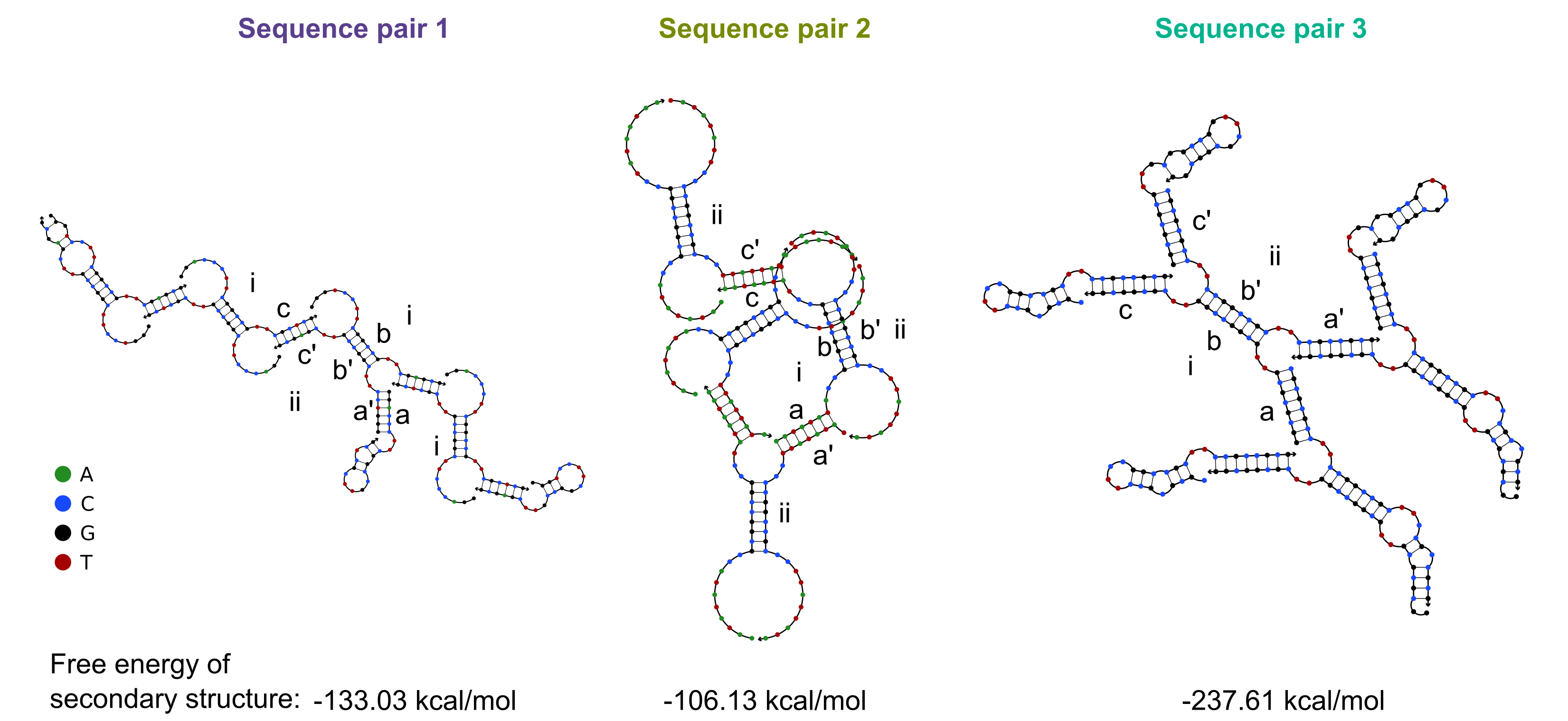}
    \caption{\textbf{Folding predictions of NUPACK.} The sequence pairs 1, 2 and 3 were analysed using the online NUPACK tool, which is based on a binding energy data set \cite{santalucia1998unified}. Prediction settings were: 15°C, 125mM NaCl, 10mM MgCl2 and 2µM per strand, secondary structure of up to 10 strands. For each system, the strand design leads to the binding pattern necessary to produce a network, in which each strand bound to the first increases the total number of possible binding sites by 1. Labeled parts of the sequences (e.g. i, ii, a, b' etc.) indicate the sequence segments as displayed in Fig.~\ref{fig:seq_design}.}
    \label{fig:NUPACK}
\end{figure*}

The sequence pairs were designed to be as short as possible, while still forming a branching network upon hybridization. The goal was to design a sequence pair composed of two sequences containing at least 3 unique binding regions (for the first strand a, b, and c  and for the second strand a', b' and c', respectively). See Fig.~\ref{fig:seq_design} for the binding scheme), such that every strand bound to the first one results in an additional binding site available. This yields $b = 3 + n$ vacant binding sites with $n$ being the number of strands bound to the first one. Therefore, the strand network will grow faster the more strands are already bound.
For the design, the Nupack Tool 3.2.2~\cite{zadeh2011nupack} was used (see Fig.~\ref{fig:nupackcode} for an exemplary code).  While keeping the center part of each sequence restricted to G or C only, the outer sections ("arms") were varied, to either consist of all four bases (Sequence pair 1), A or T only (Sequence pair 2) or GC only (Sequence pair 3). 

The resulting output sequences were iteratively mutated afterwards. Upon each mutation of the strands, the sequences were analysed using the Analysis segment of the online NUPACK tool in order to check their ability to form a network (Fig.~\ref{fig:NUPACK}). Specifically, the presence of secondary structures that included strands bound to three other strands were considered indicative of network formation.  Spacers of two bases were inserted between the three segments of each sequence. These were chosen to be either TT (Sequence pair 1 and 3) or CC (Sequence pair 2) in order to not have complementarity with the arms and reduce their participation in the overall secondary structure. Inspired by Ref.~\cite{Nguyen2017}, these spacers allow for more flexibility of the segments by minimizing angular constraints.

\begin{figure}[h]
    \centering
\begin{Verbatim}
temperature[C] = 37.0
material = dna

# domains
domain a1 = S2
domain a2 = W2
domain a3 = S2
domain b1 = S2
domain b2 = W2
domain b3 = S2
domain c1 = S2
domain c2 = W2
domain c3 = S2
domain s = T2

# strands
strand s1 = a1 a2 a3 s b1 b2 b3 s c1 c2 c3
strand s2 = a3* a2* a1* s b3* b2* b1* s c3* c2* c1*

# complexes
complex ca = s1 s2
complex cb = s1 s2
complex cc = s1 s2

# target structures
ca.structure = D6(U16 +) U16
cb.structure = U8 D6( U8 + U8) U8
cc.structure = U16 D6(+ U16)

# tubes
tube tub = ca cb cc
tub.ca.conc[M] = 1e-6
tub.cb.conc[M] = 1e-6
tub.cc.conc[M] = 1e-6

prevent = AAAA,CCCC,GGGG,UUUU,AA,GGG
stop[%] = 10
\end{Verbatim}
    \caption{\textbf{Exemplary Nupack design code} For DNA at a temperature of 37°C, two strands were designed by segmentation into smaller domains (a, b and c). In the \# domains section, the a, b and c are split into even smaller domains of similar nucleotide content  (S being G or C, and W being A or T). In this example code, TT spacers were defined via "s = T2". In "\# strands", the sequences of the system are defined, reading from 5' to 3'. The * denotes complementarity. Hybridization constraints are shown in "\# target structures", where the 3 possible binding interactions from "\# complexes" between strand 1 and strand 2 are defined further. "D6U16" for example denotes a 6 base pair region followed by a 16 unpaired region. The three target structures are also depicted in Fig.~\ref{fig:seq_design}{a}. Preventing specific sequence-patterns, such as AAAA or GGGG, above a certain relative amount, can be useful to avoid for example G-quadruplex structures or unwanted stacking.}
    \label{fig:nupackcode}
\end{figure}

\clearpage
\section{Sequence pairs 2 and 3}
\label{app:seq_2_3}
\setcounter{figure}{0}
\setcounter{table}{0}
Sequence pairs 2 and 3 did not form a condensed phase. Note that these have longer arms (a,a') and  (c,c') than sequence pair 1, which should lead to stronger binding interactions through base-pairing, and therefore more stable networks. A possible explanation for this would be that sequence pairs 2 and 3 have arms composed of only two nucleotides (GC or AT, respectively). This leads to an increased number of possible base-pairing interactions, through partial binding of the arms, due to the reduced alphabet. Most of these would be non-specific, i.e. would lead to alternative secondary structures that could inhibit network formation and suppress the growth of condensed DNA nuclei.

\section{Sedimentation analysis}
\setcounter{figure}{0}
\setcounter{table}{0}

\begin{figure*}[h]
    \centering
    \includegraphics{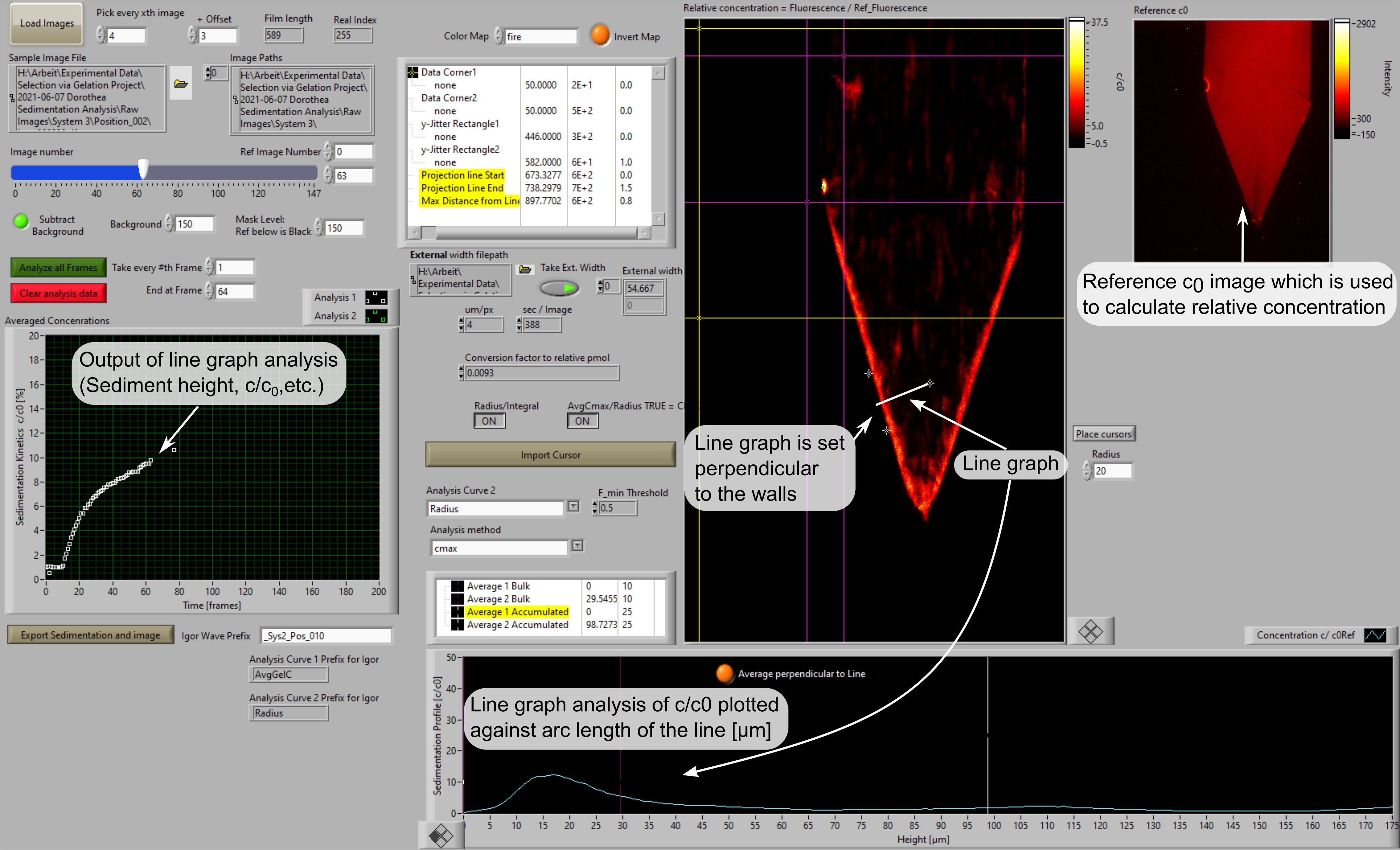}
    \caption{\textbf{Screenshot of the self-written sedimentation analysis software in LabVIEW.} A stack of images is analysed by defining a line perpendicular to the chamber wall ($x$-direction), along which the relative concentration ${c}/{c_0}$ is plotted for each image.}
    \label{fig:SedAnalysis}
\end{figure*}

In order to perform the sedimentation analysis, the time lapse stack of micrographs for each sample is loaded into a self-written Labview script (Fig.~\ref{fig:SedAnalysis}). 
Since SyBR Green I fluorescent intensity scales linearly with the amount of dsDNA in solution, it can be used for quantification~\cite{zipper2004investigations}. The first micrograph is acquired before the sedimentation starts. It is used  as a reference image after the temperature has reached 15°C. All the remaining micrographs are divided by this one to obtain relative concentration ${c}/{c_0}$. 

When a sediment is present, the relative concentration through the sediment is obtained by measuring the concentration along a defined line perpendicular to the wall of the well, which we parametruze by $x$. The maximum of relative concentration occurs at the center of the sediment. Sediment height is determined along the $x$-direction from the center concentration $c_\text{max}$ until the value has reached a $0.5 \, c_\text{max}$. The sediment height is then the distance where $c> 0.5 \, c_\text{max}$. The relative average sediment concentration is calculated by averaging over all points for which $c> 0.5 \, c_\text{max}$. 

Using the sediment height $h_\text{sed}$ and average relative concentration $\bar{c}/{c_0}$, we calculated the total amount of sedimented material $N_\text{sed}$ through Eq.~\eqref{eq:total_moles}, where $c_0$ is the initial concentration, $L$ is the length of the chamber and $d$ is the depth of the chamber:
\begin{equation}
N_\text{sed} = \left(\frac{\bar c}{c_0}\right) c_0 \,  h_\text{sed} \,   L \, d \, .
\label{eq:total_moles}
\end{equation}

\label{app:SedAnalys}

\section{Melting curves}
\label{app:melt}
\setcounter{figure}{0}   
Thermal melting curves were measured using SYBR Green I fluorescence in a thermal cycler with read-out. The samples were mixed the same way as for the sedimentation experiments: 25µM of each DNA strand, 5x Sybr Green I, 10mM Tris Buffer pH 7, 125mM NaCl and 1mM MgCl$_2$. 
Three independent mixtures were pipetted to provide independent triplicates. Additionally, a reference mixture with Sybr Green, buffer and salts was also measured to correct the data for background signals. 

The analysis of the melting curves was done with a self-written Labview script and based on the baseline adjustment described in \cite{mergny2003}. First the signal from the background fluorescence is substracted from the  fluorescence of the sample. Afterwards, the lower and higher baseline (linear) functions are determined and used for the baseline adjustment. These correspond to fully bound and fully unbound duplex states, respectively. 

\begin{figure*}[h]
    \centering
    \includegraphics{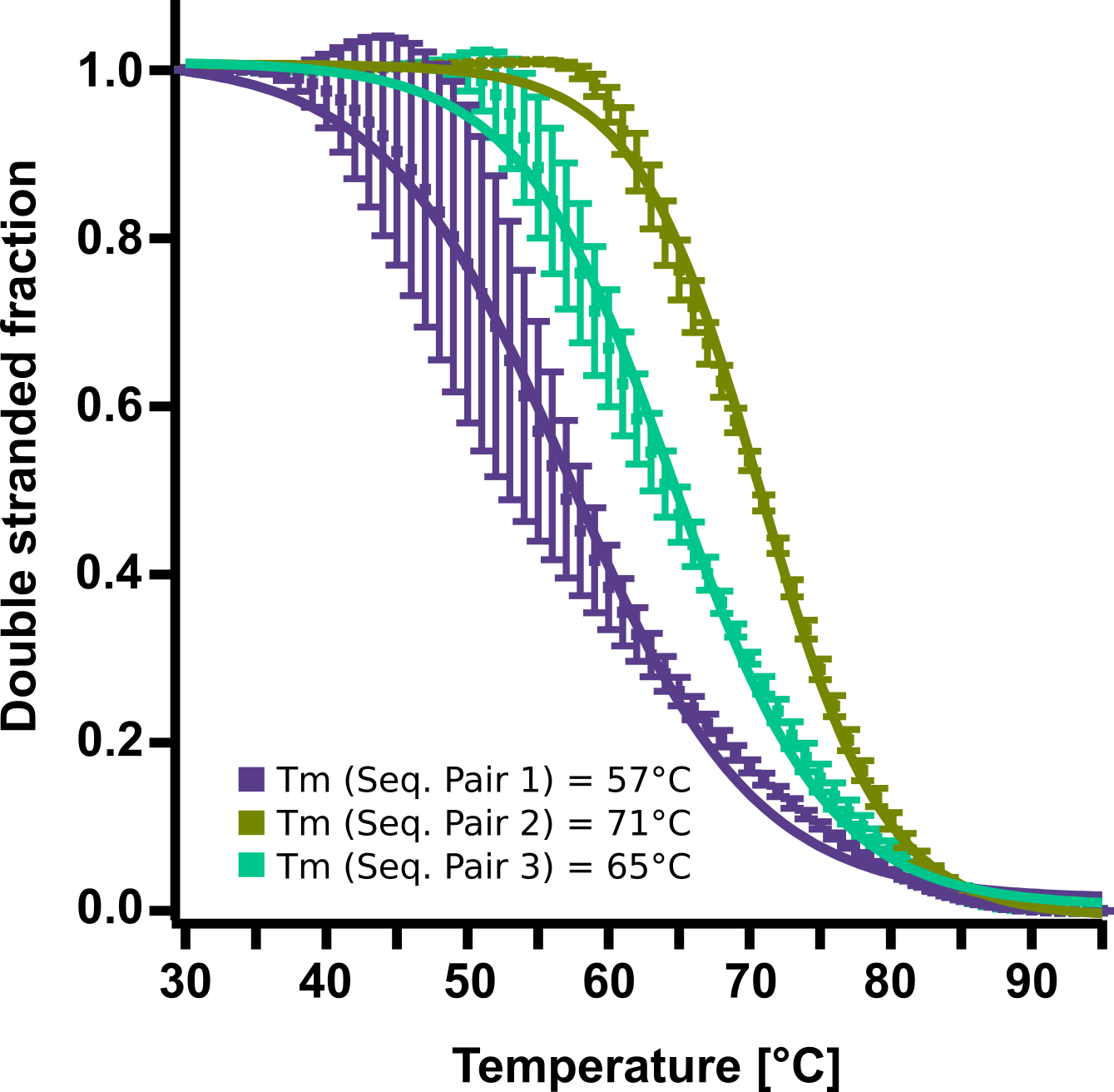}
    \caption{\textbf{Melting curves of sequence pair 1, 2 and 3.} Each point corresponds to the average between independent triplicates. Error bars depict one standard deviation of the mean. The mixtures contained both strands of each system at 25µM each DNA strand, 10mM Tris Buffer pH 7, 125mM NaCl and 10mM MgCl2. SYBR Green I concentration was 5X.}
    \label{fig:melting_curves}
\end{figure*}

The corrected data are then exported to ``Igor Pro 6.37'' and fitted with a sigmoidal function, where the midpoint fitting parameter corresponds to the melting temperature $T_m$ (see Fig.~\ref{fig:melting_curves}). The $T_m$ of all the sequence pairs was below the denaturing temperature used in the thermal protocol (95°C, see Fig.~\ref{fig:exp_cycles}a), so we ensure the mixture is homogeneous before triggering phase separation through cooling. 

\section{Mass-balance over refeeding cycles}
\label{app:mass-balance}

\begin{figure*}[h]
    \centering
    \includegraphics{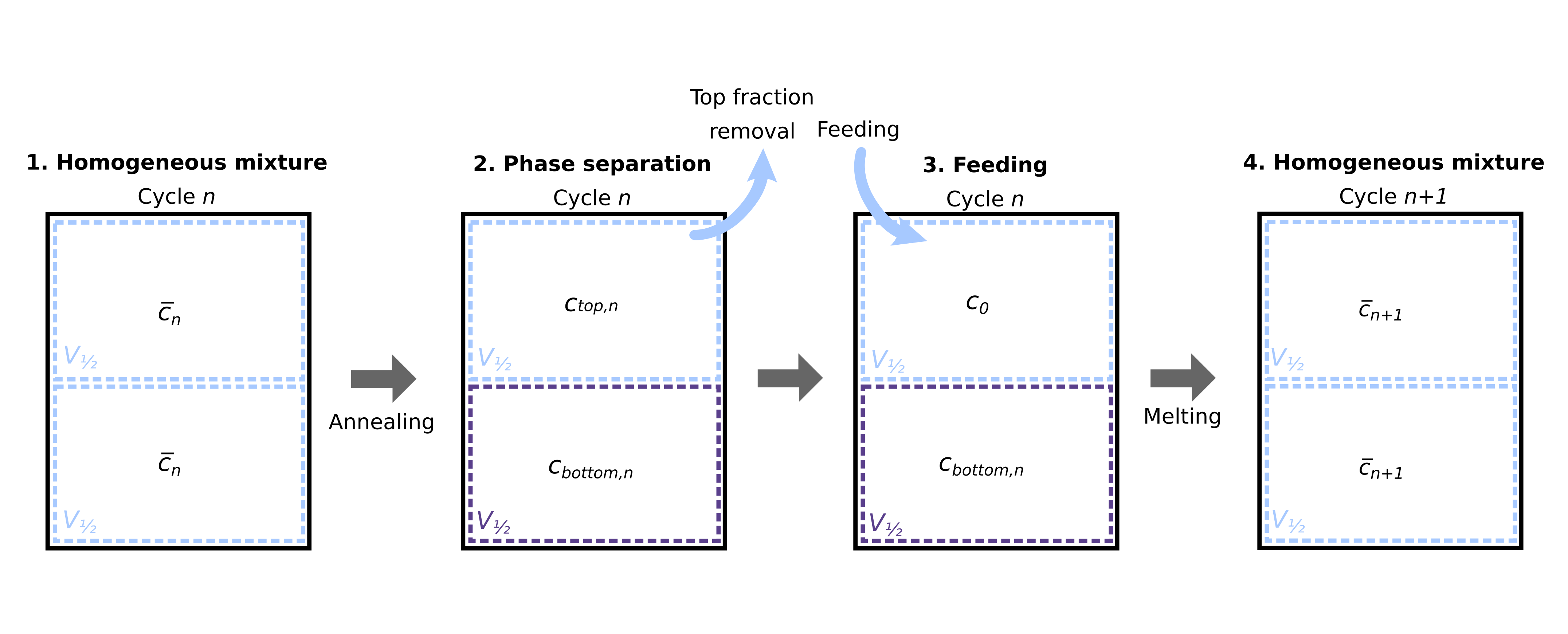}
    \caption{\textbf{Schematic of a feeding cycle and the corresponding concentrations for top and bottom volume fractions.} }
    \label{fig:SI_Massbalance}
\end{figure*}

We use mass balance to determine the concentration of the bottom fraction for each cycle $n$. 
Using the concentration of the removed top fraction through HPLC analysis, and the concentration of the initial pool $c_0$, we calculate all the intermediate bottom fractions as plotted in Fig.~\ref{fig:exp_cycles}b. 
At step 1, the system phase separates which conserves the total mass (Fig.~\ref{fig:exp_cycles}a):
\begin{align}
\label{eq:mat_cons}
&\bar{c}_n \, 2 V_\frac{1}{2} = c_{\text{top},n} \, V_\frac{1}{2} + c_{\text{bottom},n} \, V_\frac{1}{2} \, ,  \qquad \text{and thus} \quad c_{\text{bottom},n} = 2 \bar{c}_n - c_{\text{top},n} \, ,
\end{align}
where $V_\frac{1}{2}= V /2$ denotes half the total volume ($V$ is the total volume of the system), $c_{\text{top},n}$ and $c_{\text{bottom},n}$ the concentrations of material in the top and bottom fraction in the $n$-th cycle, respectively. Note that the bottom fraction includes the sedimented dense phase. Moreover, $\bar{c}_n$ describes the concentration in the whole volume during the $n$-th cycle, which is conserved during phase separation (see Fig.~\ref{fig:SI_Massbalance}).

At step 2 the top half of the system is removed and fed at step 3 by the pool of concentration ${c}_\text{pool}$ (Fig.~\ref{fig:exp_cycles}a). The corresponding mass balance  reads: 
\begin{align}
\label{eq:mat_cons2}
&{c}_\text{pool} \, V_\frac{1}{2} + c_{\text{bottom},n} \, V_\frac{1}{2} =  \bar{c}_{n+1} \, 2 V \,  .
\end{align}
Using Eq.~\eqref{eq:mat_cons} to substitute Eq.~\eqref{eq:mat_cons2},  $\bar{c}_{n+1}$ can be written as a function of $c_{\text{top},n}$, $c_n$ and ${c}_\text{pool}$. This relation can be used to obtain an equation for $c_{\text{bottom}, n+1}$, through substitution in Eq.~\eqref{eq:mat_cons}:
\begin{align}
\bar{c}_{n+1} &= \bar{c}_n + \frac{{c}_\text{pool} - c_{\text{top},n}}{2} \, ,\\
c_{\text{bottom},n+1} &= 2\bar{c}_n + {c}_\text{pool} - c_{\text{top},n} - c_{\text{top},n+1} \, .
\label{eq:mat_cons3}
\end{align}
Since for $n=0$, $\bar{c}_n={c}_\text{pool}$ and by  measuring the top fraction concentrations $c_{\text{top},n+1}$, we can recursively calculate all the bottom fraction concentrations using Eq.~\eqref{eq:mat_cons3}.

\section{Finite element calculations}
\label{app:comsol}
\setcounter{figure}{0}
\setcounter{table}{0}

The experimental chamber was designed using an Autodesk CAD software (Inventor 2019). The 2D drawing was then exported to COMSOL Multiphysics 5.4 using the Inventor live-link plugin. The geometry is therefore matching the experimental chamber in the two dimensions $x$ and $y$. Since the flows and the sedimentation of material occur only in $x$- and $y$-direction, omitting the $z$-axis allowed us to effectively screen parameters focusing on the key dynamics of the system. 

The inlet and outlet of the well were emulated defining a constant normal inflow speed of 2µm/s as well as an outlet with a pressure boundary condition. Assuming a constant 20°C across the whole geometry, stationary laminar flow was solved with Navier-Stokes equations for conservation of momentum and continuity equation for conservation of mass.
Since the flow speed of 2µm/s is very low (Mach number M < 0.3), the flow can be considered as incompressible. Thus, the density is assumed to be constant and the continuity equation reduces to the condition:
\begin{equation}
\nabla \cdot \vect{u}=\vect{0} \, .
\end{equation}
with $\rho$ denoting the mass density and $\vect{u}$ the velocity field. The Navier-Stokes equation then reduces to 
\begin{equation}
\rho(\vect{u}\cdot\nabla)\vect{u} = \nabla \cdot [-p \vb*{I} + \eta(\nabla\vect{u}+(\nabla \vect{u})^T] + \vb*{F} = \vb*{0} \, ,
\end{equation}
with $p$ being pressure, $\vb*{I}$ the unity tensor, $\eta$ the fluid dynamic viscosity and $\vb*{F}$ the external forces applied to the liquid. The reference pressure was set to 1[atm], the reference temperature was 20°C and all surfaces are described as non-slip boundary conditions. 

\begin{figure*}
\includegraphics{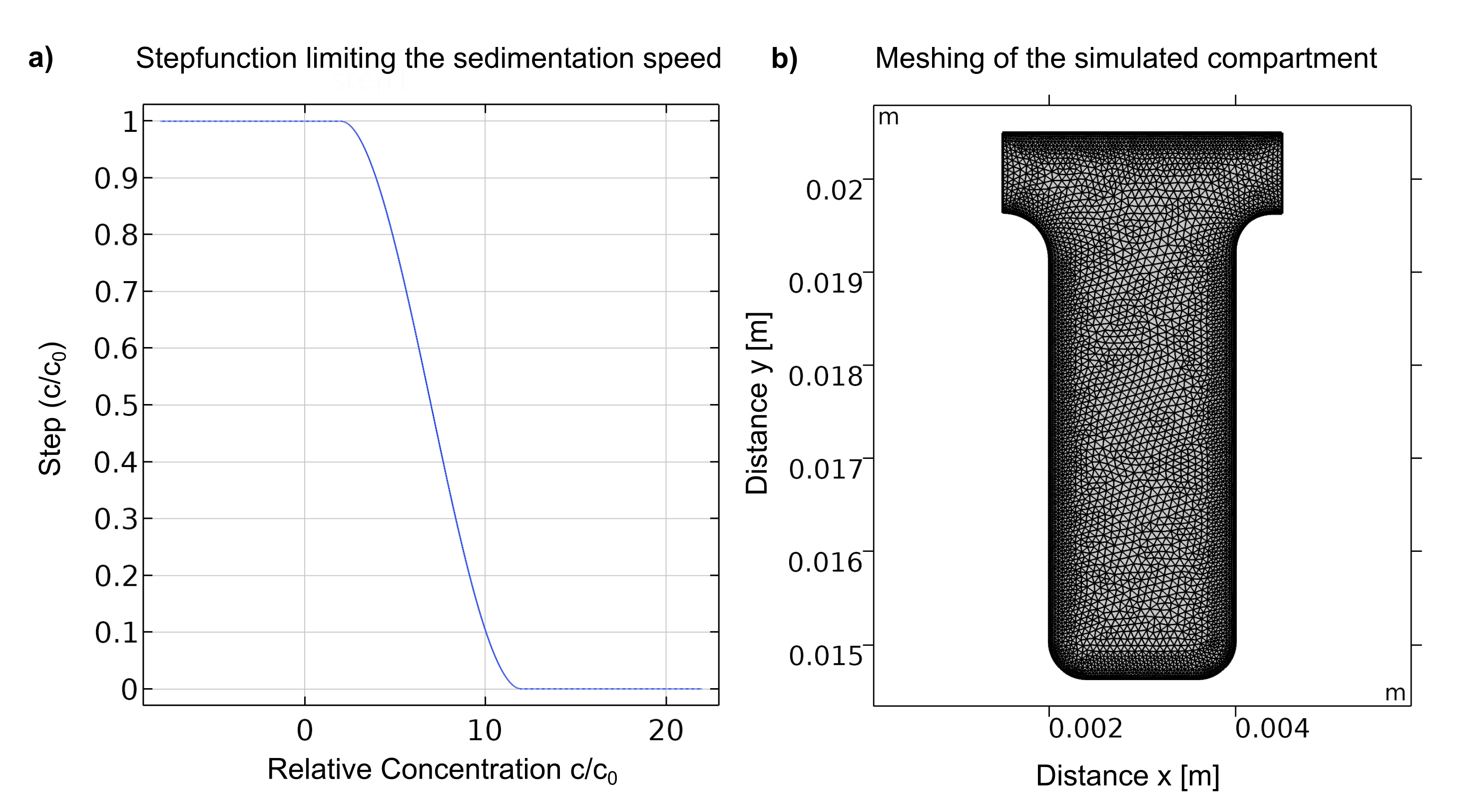}
\caption{\textbf{Sedimentation speed stepfunction and compartment meshing}. \textbf{a} Stepfunction dependent on local relative concentration $c/c_0$. This function is multiplied with the sedimentation velocity in each respective mesh triangular. Thus, molecule sedimentation arrests as they reach the  bottom of the well. 
\textbf{b} Depiction of the mesh-size used for the simulation. The surfaces required more refined meshing to avoid problems related to boundary conditions. 
}
\label{fig:ComsolDetails}
\end{figure*}

Assuming the laminar flow to be stationary, we described the transport of a diluted species $c_i$ by combining convection and diffusion effects:
\begin{equation}
\frac{\partial c_{i}}{\partial t} +  \nabla (-D_i\nabla c_i + \vect{u}c_i)= 0 \, ,
\end{equation}
where $D_i$ denotes the diffusion coefficient of species $i$.
Sedimentation of oligomers was simulated assuming a downwards oriented flow speed $v_{sed}$ affecting diluted species only. With increasing local relative concentration $\frac{c}{c_0}$, the flow speed decreases from 0.1µm/s to 0, using an inbuilt smoothened heavyside step function, which is shown in Fig.~\ref{fig:ComsolDetails}. The $y$-component of the velocity field $\vect{u}$ used to describe the convective movement of diluted species $c_i$ then reads: 
\begin{equation}
{u}_y = v- {v}_{sed} \quad \text{with}\quad {v}_{sed} = v_{0}\cdot step\left(\frac{c_i}{c_0}\right) \, .
\end{equation}
Here, ${v}_{sed}$  ensures that molecules cannot sediment into an area, where the local concentration has already reached ${c_i}/{c_0} = 10$. This maximum density was estimated using the fluorescence data from the experiments. Choosing the center of the step function to be at ${c}/{c_0} = 7$ gave best results matching the experiments. The sedimentation speed of maximum 0.1 µm/s was chosen through observation of the distance traveled in between images for the very first visible condensed DNA aggregates. The aggregates sediment quicker, the larger they become. These kinetics were not simulated. Rather, the system was treated as if it was filled with those small aggregates from the beginning, sedimenting with $v_\text{sed}$. Meshing was done using an automatic finer physics-controlled mesh (see Fig.~\ref{fig:ComsolDetails}. After setting the parameters as above, a complete numerical solution could be found. A complete set of all input parameters can be found in Table~\ref{tab:Comsol_Parameters}. 

\begin{table}[]
    \centering
    \begin{tabular}{l|l|l}
    \textbf{Parameter} & \textbf{Value} & \textbf{Description} \\
      $v_\text{in} $  &  2 $ [\mu m/s]$  & Inflow velocity \\
      $D$   &  5 $ [\mu m^2/s]$ & Diffusion constant of species c \\
      $v_{0}$ & 0.1$ [\mu m/s]$ & Sedimentation velocity \\
      $c_{0}$ & 25 $ [\mu M]$ & Initial concentration of species c \\
      well widht & 2 $ [mm]$ & Width of the simulated pore \\
      well height & 5 $[mm]$ & Height of the simulated pore \\
      sed cutoff & 7 & Position of the center of the stepfunction \\
      sed smoothing & 10 & width of the stepfunction 
    \end{tabular}
    \caption{\textbf{Parameters used for the numerical calcualtion for the system with continuous feeding flow.} Final set of parameters usd to simulate the flowthrough experiment from Fig.~\ref{fig:flow}. Water-specific parameters such as dynamic viscosity or density were taken from inbuilt features of COMSOL Multiphysics 5.4..}
    \label{tab:Comsol_Parameters}
\end{table}

\section{Experimental setup}
\label{app:setup}
\setcounter{figure}{0}   
The self-built experimental setup used to conduct and image sedimentation experiments is composed of:
\begin{enumerate}
    \item Temperature control and chamber mount (Fig.~\ref{fig:wells} and Fig.~\ref{fig:flow2})
    \item Fluorescence microscope (Fig.~\ref{fig:SISetup})

\end{enumerate}

\begin{figure*}[h]
    \centering
    \includegraphics{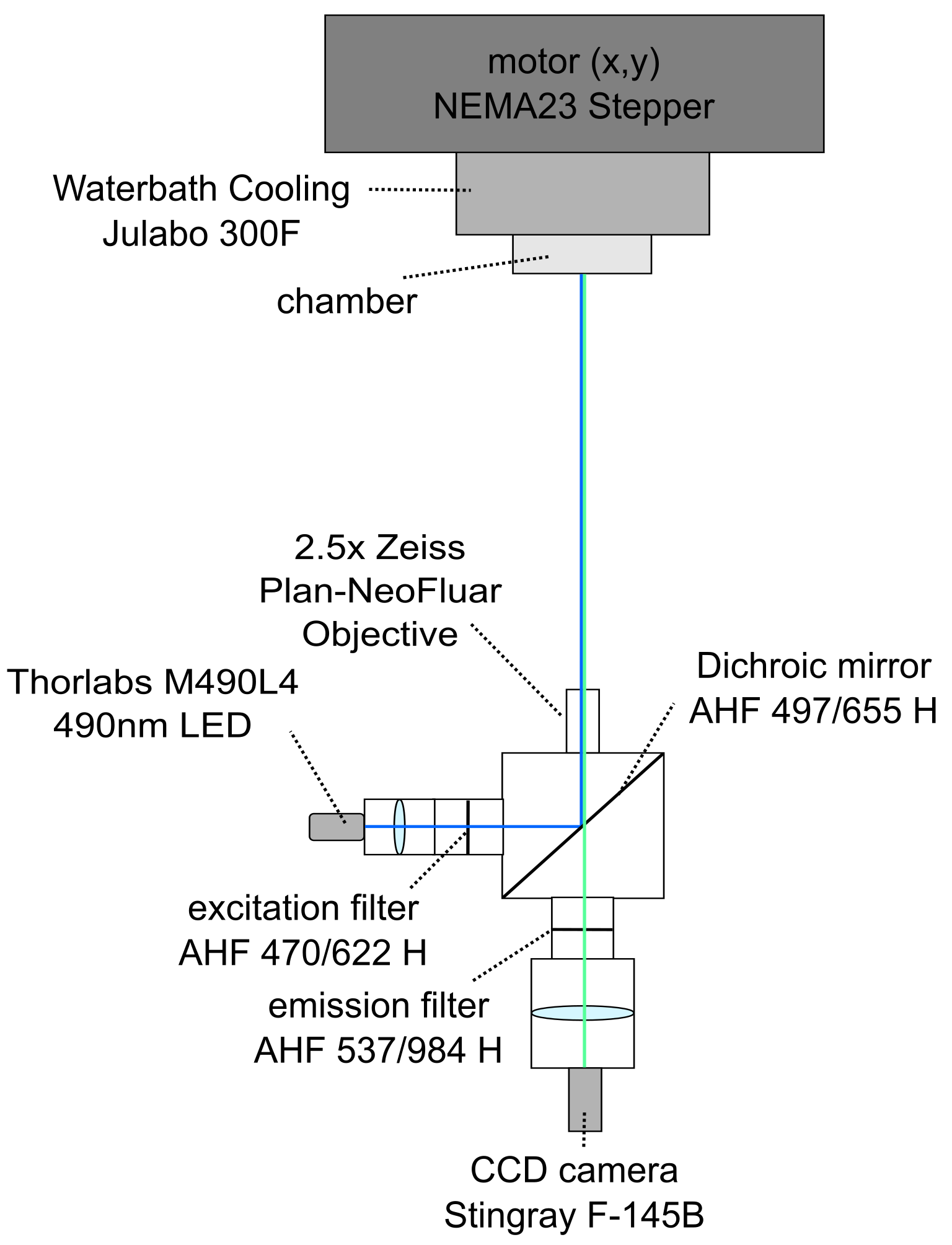}
    \caption{\textbf{Schematic drawing of the measurement setup used to take fluorescent images for all experiments} Motors, LED's as well as the camera were controlled using a self-written LABview software. The setup was surrounded by black curtains during measurements.}
    \label{fig:SISetup}
\end{figure*}

\begin{figure*}[h]
\includegraphics{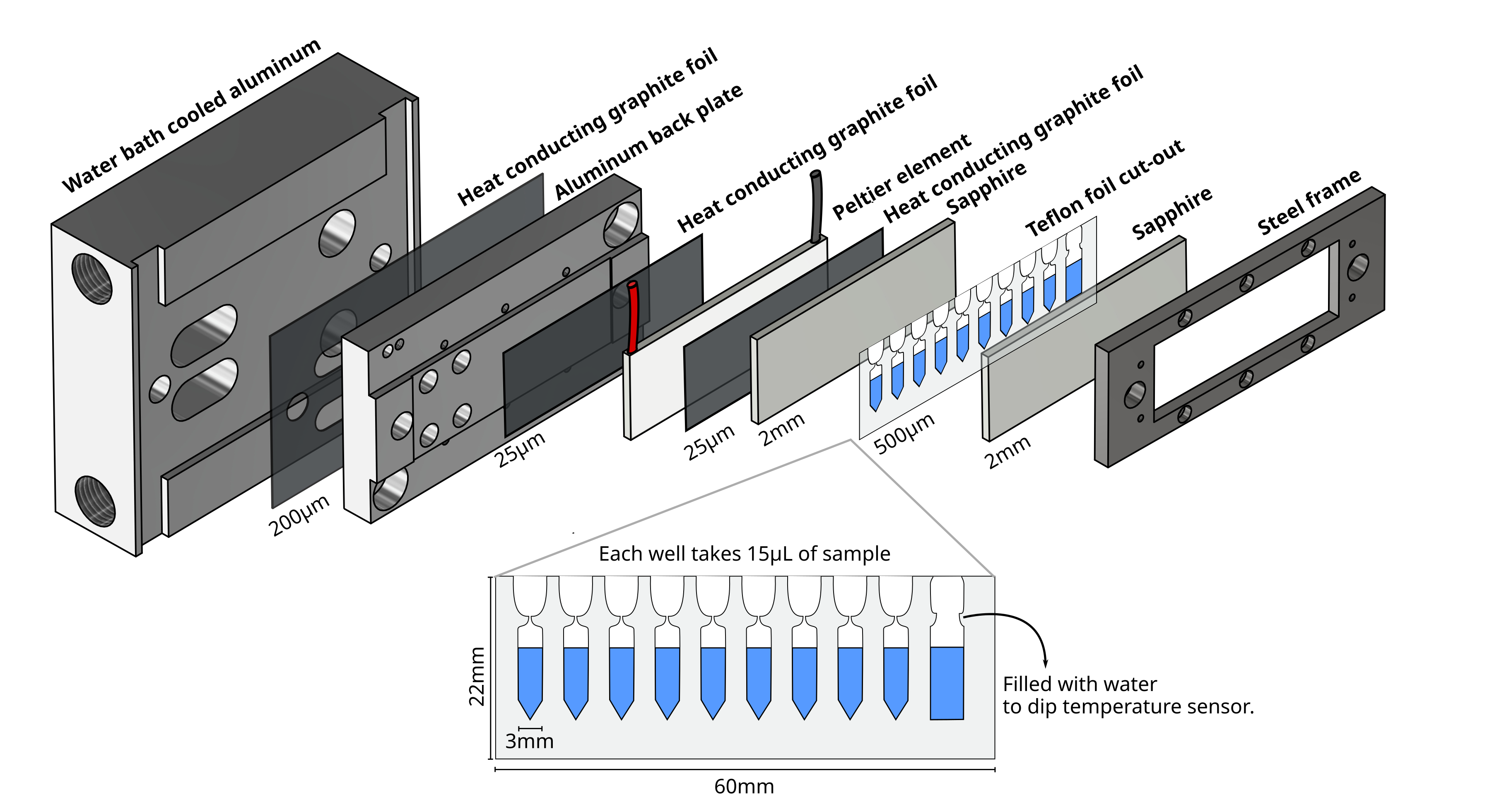}
\caption{\textbf{Schematic of sedimentation wells.} \textbf{a} The chamber is mounted from left to right. The temperature is controlled with the Peltier element, heat conducting graphite foil is used between all the parts to enhance heat conductivity. The waterbath is used to dissipate the heat generated by the rear panel of the Peltier element back. The microfluidic Teflon cut-out is sandwiched between two sapphires which  provide high heat conductivity when coupled to visible light transparency. The thickness is shown below each of the relevant parts. 
\textbf{b} Teflon cut-out has 9 each 3mm-wide wells and a temperature sensor  is connected to each well. The shape of the chamber was designed to prevent the evaporation of the sample upon heating. The wide area above the narrow bottleneck is filled with parafilm, which seals the well. 
}
\label{fig:wells}
\end{figure*}

\begin{figure*}[h]
\includegraphics{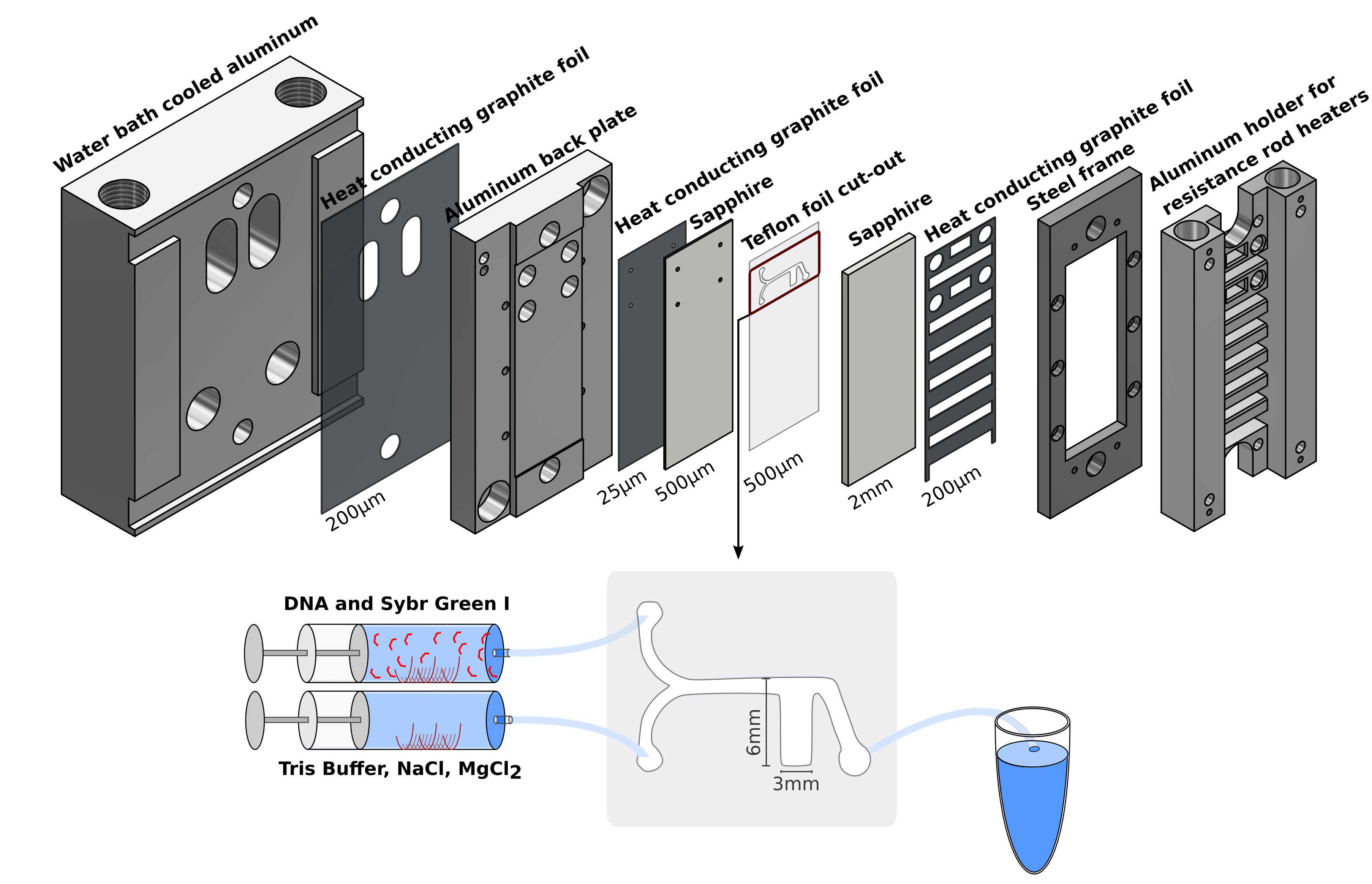}
\caption{\textbf{Schematic of the chamber used for the continuous feeding experiments.} 
\textbf{a} The chamber is mounted from left to right. The temperature is controlled with the water bath (set to 15°C), heat conducting graphite foil is used between all the parts to enhance heat conductivity. The microfluidic Teflon cut-out is sandwiched between two sapphires, used for their high heat conductivity coupled to visible light transparency. The thickness is written below each of the relevant parts. 
\textbf{b} Two 100µL syringes are used to inject the sample in the microfluidic well. DNA and Sybr Green I was injected through a separate inlet than the Tris Buffer and salts to prevent sedimentation to happen in the syringes and tubings. 
}
\label{fig:flow2}
\end{figure*}

\section{Supplementary videos}
\label{app:videos}
\setcounter{figure}{0}
\setcounter{table}{0}

\begin{itemize}
    \item Video 1:`Video1.avi'' Timelapse video of Sequence pair 1 condensing and sedimenting inside the chamber. Chamber depth is 500$\mu$m. Temperature profile as described in methods. Contrast and brightness have been modified for qualitative reasons, the analysis of the readout was performed using the raw images.
    Sample contained 25$\mu$M of sequence pair 1 10mM TRIS pH 7, 5X Sybr Green I, 125mM NaCl and 10mM MgCl$_2$.
    \item Video 2:`Video2.avi'' Timelapse video of sequence pair 1 and 2 condensing and sedimenting inside the chamber. Chamber depth is 500$\mu$m. Temperature profile as described in methods. Contrast and brightness have been modified for qualitative reasons, the analysis of the readout was performed using the raw images.
    Sample contained 25$\mu$M of both sequence pairs 10mM TRIS pH 7, 5X Sybr Green I, 125mM NaCl and 10mM MgCl$_2$.
    \item Video 3:`Video3.avi'' Timelapse Video of the flowthrough experiment shown in Fig.~\ref{fig:flow}.  Sample contained 25$\mu$M of sequence pair 1 10mM TRIS pH 7, 5X Sybr Green I, 125mM NaCl and 10mM MgCl$_2$. Chamber depth is 500$\mu$m. Contrast and brightness have been modified for qualitative reasons, the analysis of the readout was performed using the raw images.
\end{itemize}


\end{document}

%% file: preamble.tex
\usepackage{mathtools}
\usepackage{physics}
\usepackage{xcolor}
\usepackage{graphicx}
\usepackage{adjustbox}
\usepackage{placeins}
\usepackage[T1]{fontenc}
\usepackage{lipsum}
\usepackage{csquotes}

\makeatletter
\renewcommand*{\fnum@figure}{{\normalfont\bfseries \figurename~\thefigure}}
\renewcommand*{\@caption@fignum@sep}{\textbf{: }}
\makeatother

\usepackage{chemfig}
